\documentclass[11pt]{article}
\usepackage{geometry} 
\usepackage[usenames,dvipsnames]{color}
\usepackage{amssymb}
\usepackage{amsmath}
\usepackage{amsfonts}
\usepackage{bbm}
\usepackage{cancel}
\usepackage{cite}

\usepackage{amsthm}
\theoremstyle{definition} 

\newtheorem{definition}{Definition}
\theoremstyle{theorem} 

\newtheorem{proposition}{Proposition}

\usepackage{hyperref}
\hypersetup{
   colorlinks=true,         
   linkcolor=Magenta,          
   citecolor=red,        
   urlcolor=Red            
}

\def\be{\begin{equation}}
\def\ee{\end{equation}}
\def\bea{\begin{eqnarray}}
\def\eea{\end{eqnarray}}

\def\1{\'{\i}}                           

\def\>#1{{\mathbf#1}}

\def\conm#1#2{\left[ #1,#2 \right]}  
\def\pois#1#2{\left\{ #1,#2 \right\}}

 \def\a{{\alpha}}  
 \def\b{{\beta}}  
 \def\c{{\gamma}}  

 \def\h{{\hat\xi}}  
\def\t{{\hat x}}

\begin{document}

\title{
Coisotropic Lie bialgebras and complementary dual Poisson homogeneous spaces
}

\author{Angel Ballesteros$^1$, Ivan Gutierrez-Sagredo$^{1,2}$ and Flavio Mercati$^1$
\vspace{12pt}
\\
$^1$Departamento de F\'isica, Universidad de Burgos, \\
E-09001 Burgos, Spain 
\vspace{12pt} \\
$^2$Departamento de Matem\'aticas y Computaci\'on, \\
E-09001 Burgos, Spain 
\vspace{12pt}
\\
E-mail: \href{mailto:angelb@ubu.es}{angelb@ubu.es}; \href{mailto:igsagredo@ubu.es}{igsagredo@ubu.es}; \href{mailto:fmercati@ubu.es}{fmercati@ubu.es}}

\maketitle

\begin{abstract}

Quantum homogeneous spaces are noncommutative spaces with quantum group covariance. Their semiclassical counterparts are Poisson homogeneous spaces, which are quotient manifolds of Lie groups $M=G/H$ equipped with an additional Poisson structure $\pi$ which is compatible with a Poisson-Lie structure $\Pi$ on $G$. Since the infinitesimal version of $\Pi$ defines a unique Lie bialgebra structure $\delta$ on the Lie algebra $\frak g=\mbox{Lie}(G)$, we exploit the idea of Lie bialgebra duality in order to study the notion of complementary dual homogeneous space $M^\perp=G^\ast/H^\perp$ of a given homogeneous space $M$ with respect to a coisotropic Lie bialgebra. Then, by considering the natural notions of reductive and symmetric homogeneous spaces, we extend these concepts to $M^\perp$ thus showing that an even richer duality framework between $M$ and $M^\perp$ arises from them. In order to analyse physical implications of these notions, the case of $M$ being a Minkowski or (Anti-) de Sitter Poisson homogeneous spacetime is fully studied, and the corresponding complementary dual reductive and symmetric spaces $M^\perp$ are explicitly constructed in the case of the well-known $\kappa$-deformation, where the cosmological constant $\Lambda$ is introduced as an explicit parameter in order to describe all Lorentzian spaces simultaneously. In particular, the fact that $M^\perp$ is a reductive space is shown to provide a natural condition for the representation theory of the quantum analogue of $M$ that ensures the existence of physically meaningful uncertainty relations between the noncommutative spacetime coordinates.  Finally, despite these dual spaces $M^\perp$ are not endowed in general with a $G^\ast$-invariant metric, we show that their geometry can be described by making use of $K$-structures.

 \end{abstract}

\noindent
KEYWORDS:  Homogeneous spaces, reductive and symmetric spaces, Lie bialgebras, Poisson--Lie groups, non-commutative  spacetimes, uncertainty relations


\bigskip

\tableofcontents

\section{\bf Introduction}

General Relativity can be understood as a theory of spacetime geometry, and therefore it is expected that (at least some) of the essential features of Quantum Gravity could be interpreted in terms of certain `quantum geometries' in which Planck scale effects can be appropriately described and predicted. One of the most relevant qualitative effects that arise from different approaches to Quantum Gravity is the existence of nonvanishing uncertainty relations among the observables describing spacetime coordinates (see for instance~\cite{Maggiore1993algebraicgup,DFR1994,Garay1995,Szabo2003} and references therein), which implies that the latter are described through a noncommutative spacetime algebra. In this context, quantum homogeneus spaces (see~\cite{DijkhuizenKoornwinder1994} for a comprehensive review) arise as a sound possibility in order to model such `quantum spacetimes', since they are -by construction- covariant under quantum group (co)actions~\cite{ChariPressley1994,Majid1995book}. Therefore, it is interesting to construct quantum analogues of the isometry groups of maximally symmetric Lorentzian spacetimes (Minkowski and (Anti-) de Sitter) together with their associated quantum homogeneous spacetimes  (see, for instance,~\cite{Majid1988,PW1990,LRNT1991,MR1994,Zakrzewski1994poincare,BHOS1994global,Podles1995, BHOS1995nullplane,Ciccoli1996qplanes, BCGST1996,Brzezinski1996, BRH2003minkowskian,BLT2016unified, BLT2016unifiedaddendum, MS2018constraints, BM2018extended,BGH2019kappaAdS3+1}).

On the other hand, it is worth stressing that quantum homogeneous spaces are quantizations (see~\cite{DijkhuizenKoornwinder1994, Ciccoli1996qplanes,Karolinsky2005} for technical details) of Poisson homogeneous spaces (hereafter PHS), which are covariant under Poisson-Lie group actions~\cite{Drinfeld1993}, and the quantization of the latter provides  the corresponding quantum group symmetry. As a consequence, the study and explicit construction of Poisson homogeneous spacetimes has proven fruitful in order to construct quantum homogeneous spacetimes and, in general, noncommutative spaces with quantum group invariance (see~\cite{Zakrzewski1991,Zakrzewski1994homogeneous, Ciccoli1996qplanes, Reyman1996, EK1998, CG2006qd, BMN2017homogeneous, BGH2019worldlinesplb,BGGH2020kappanewtoncarroll} and references therein).

In order to endow a given homogeneous space $M=G/H$, in which $H$ is the isotropy subgroup of the origin, with a Poisson homogeneous structure  $\pi$, we start from a Poisson-Lie (PL) structure $\Pi$ on $G$ and we impose that the homogeneous space action $\rhd: G \times M\rightarrow M$ is a Poisson map. Moreover, the Poisson homogeneous structure $\pi$ on $M$ can be obtained as the canonical projection of $\Pi$ provided that the unique Lie bialgebra $(\frak g,\delta)$ associated to $(G,\Pi)$ is $(\mathfrak h,\mathfrak t)$-coisotropic with respect to the Lie algebra $\frak h$ of $H$, namely, if the cocommutator map is of the form
\be
\delta(\frak h) \subseteq  \frak h \wedge \frak g\, ,
\label{coisotra}
\ee
where $\mathfrak t \subset \mathfrak g$ is the orthogonal complement of $\mathfrak h$.
We remark that this condition is not invariant when a conjugate subgroup $H'$ is considered \cite{Ciccoli1996qplanes,Karolinsky1995,Karolinsky1996,Karolinsky1997,Karolinsky1998,Lu2000}. In this paper we will restrict our study to the case of the so-called pointed Poisson homogeneous spaces, in which the isotropy subgroup $H$ (and hence the origin of the homogeneous space) will be fixed. This is consistent with the fact that the Poisson structure $\pi$ on $M$ on a given point depends on the coordinates of such a point.

When the coisotropy condition is analysed at the level of the dual Lie bialgebra $(\frak g^\ast,\delta^\ast)$ that is associated to the dual Poisson-Lie group $(G^\ast,\Pi^\ast)$, it is straightforward to see that~\eqref{coisotra} implies the existence of a Lie subalgebra within $\frak g^\ast$ that we will denote $\frak h^\perp$ \cite{Lu1990thesis,CG2006qd}. As a consequence, we can use the unique connected and simply-connected Lie subgroup $H^\perp$ of $G^\ast$ with Lie algebra $\frak h^\perp$ in order to construct the so-called {\it complementary dual} homogeneous space $M^\perp$ as the space of cosets $G^\ast/H^\perp$. 
Moreover, $M^\perp$ turns out to be a Poisson homogeneous space for $G^*$, and the dual Lie bialgebra $(\frak g^\ast,\delta^\ast)$ is $(\mathfrak h^\perp,\mathfrak t^\perp)$-coisotropic with respect to the Lie algebra $\mathfrak h^\perp$ of $H^\perp$, since the dual cocommutator reads
\begin{equation}
\delta^* (\mathfrak h^\perp) \subseteq \mathfrak h^\perp \wedge \mathfrak g^* . 
\end{equation}
This ensures that the Poisson homogeneous structure $\pi^\ast$ onto $M^\perp$ can be obtained as the canonical projection of $\Pi^\ast$. We stress that the initial Poisson homogeneous space $M$ and its complementary dual $M^\perp$ have in general different dimensions as well as very different geometric properties and Poisson structures.

The aim of this paper is two-fold. On one hand, we try to provide a new insight into the geometry of the dual complementary space $M^\perp$ by characterizing the conditions that the initial Lie bialgebra cocommutator $\delta$ has to fulfill in order to allow  $M^\perp$ to be either a reductive or a symmetric homogenous space. This will give rise to the so-called {\it coreductivity} and {\it cosymmetry} conditions for $\delta$. In this way, the geometry of the dual reductive space $M^\perp$ can be characterized by making use of the theory of $K$-structures \cite{chern1953Gstructures,KobayashiNomizu1969-2}, since in general the dual spaces $M^\perp$ cannot be endowed with a $G^\ast$-invariant metric. Also, by imposing that $\frak g^\ast$ is the Lie algebra of a symmetric reductive homogeneous space $M^\perp$, cosymmetry implies that the canonical connection on $M^\perp$ is torsionless and completes the duality framework between $M$ and $M^\perp$. 

On the other hand, we will show the computational usefulness of the coreductivity and cosymmetry notions in order to characterize Poisson homogeneous spacetimes through their associated Lie bialgebras. Moreover, we will provide some explicit and physically relevant examples of dual complementary spaces corresponding to Poisson homogeneous structures for the Minkowski and (Anti-) de Sitter spaces. In particular, the dual reductive homogeneous spacetimes associated to the well-known $\kappa$-deformation~\cite{LRNT1991,MR1994, Zakrzewski1994poincare,BorowiecPachol2009jordanian,BGH2019worldlinesplb,BP2014extendedkappa,LSW2015hopfalgebroids,ABP2017,BGH2019kappaAdS3+1,GutierrezSagredo2019phd,BGGH2020kappanewtoncarroll}  will be explictly constructed, and their geometric properties will be explicitly analyzed. We will also show that, after quantization, coreductivity becomes essential for the representation theory of the (linearized) quantum homogeneous space, since the fact that $\frak g^\ast$ corresponds to a reductive complementary space implies that the restriction of the representations of the full $\frak g^\ast$ onto the noncommutative space $\mathfrak h^\perp$ can be used in order to construct representations of the latter. 

The next two sections of the paper are devoted to the presentation of all the background material which will be relevant in the following. In the next section the basic elements of the theory of Poisson-Lie groups and Lie bialgebras will be sketched, together with the notion of coisotropic Poisson-Lie subgroup. Section 3 is devoted to the theory of homogeneous spaces, and the properties of the Lie algebras underlying reductive and symmetric homogeneous spaces are described. Afterwards, Poisson homogeneous spaces (PHS) are introduced, with special emphasis in the notions of pointed and coisotropic PHS. 

In section 4, given a pointed coisotropic PHS $M$, the notion of complementary dual Poisson homogeneous space $M^\perp$ will be introduced, and the duality framework between $M$ and $M^\perp$ is ellaborated, including the description of both spaces in terms of local coordinates. In section 5 the new notions of coreductivity and cosymmetry for the complementary dual PHS $M^\perp$ will be presented. Their algebraic characterization will be given in terms of suitable constraints on the cocommutator $\delta$ of the Lie bialgebra, which allow to distinguish from the very beginning whether a given coisotropic PL structure $\Pi$ allows the construction of a complementary dual space which is reductive and/or symmetric.

As we will see in Section 6, these new concepts are meaningful for a novel approach to the Lie bialgebra structures for Lorentzian Lie algebras, since coisotropy and coreductivity conditions provide strong constraints on the $r$-matrices generating (Anti) de Sitter and Poincar\'e Lie bialgebras in (2+1) and (3+1) dimensions. In particular, the well-known $\kappa$-deformation of the (Anti) de Sitter and Poincar\'e Lie algebras will be analysed from this viewpoint. It is found that while in (2+1) dimensions the $\kappa$-Lie bialgebra is coreductive for the three algebras, in (3+1) dimensions coreductivity is only admissible for the $\kappa$-Poincar\'e $r$-matrix, since the introduction of this condition turns out to be incompatible with the existence of a non-vanishing cosmological constant parameter $\Lambda$. Moreover, the corresponding dual PHS (that we shall denote as $M^\perp_\Lambda$) will be explicitly constructed. As a result, we obtain that while in (2+1) dimensions their associated Poisson structures $\pi^\ast$ turn out to be $\Lambda$-deformations of the (2+1) Lorentz Lie algebra $\frak{so}(2,1)$,  in (3+1) dimensions the dual $M^\perp_0$ of the $\kappa$-Minkowski space has a Poisson structure isomorphic to the non-deformed Lorentz Lie algebra $\frak{so}(3,1)$.

The fact that all these dual spaces  $M^\perp_\Lambda$ constructed in Section 6 cannot be endowed with a $G^\ast$-invariant metric, leads to the consideration of alternative approaches in order to unveil some of their geometric properties. With this motivation, Section 7 discusses the geometry of dual PHS from the viewpoint of $K$-structures on manifolds, which allows the definition of their curvature, torsion and Ricci tensors. In particular, it is found that the $\kappa$-Lie bialgebra in (2+1) dimensions gives rise to dual spaces $M^\perp_\Lambda$ which are always  torsionless, and whose components of the curvature tensor are proportional to $\Lambda$. In Section 8 the connection between the coreductivity condition for a given Lie bialgebra and the properties of the uncertainty relations that would arise from the noncommutative spacetime coordinates of the associated quantum homogeneous space  is discussed in terms of the representation theory of the full dual algebra $\frak g^\ast$ and its restriction to the first-order noncommutative space. Finally, a concluding Section closes the paper.



\section{Poisson-Lie groups and Lie bialgebras}

A Poisson structure on a smooth manifold $M$ can be defined in terms of the Poisson bivector $\pi \in \Gamma (TM \wedge TM)$ or equivalently, in terms of the Poisson bracket $\{ \, , \, \}: \mathcal{C}^\infty (M) \times \mathcal{C}^\infty (M) \rightarrow \mathcal{C}^\infty (M)$. Recall that the relation between them is simply $\pi (df,dg) = \{ f,g\}$ for any functions $f,g \in \mathcal{C}^\infty (M)$. In this paper we will use both notations. Among Poisson structures, those defined on a Lie group and being compatible with the group multiplication will be specially relevant for this paper. A Lie group endowed with such a Poisson structure is called a Poisson-Lie group (see~\cite{ChariPressley1994} and references therein):

\begin{definition}
A \emph{Poisson-Lie group} $(G,\Pi)$ is a Lie group $G$ endowed with a Poisson structure $\Pi$ in such a way that the multiplication $\mu : G \times G \rightarrow G$ is a Poisson map with respect to $\Pi$ on $G$ and the product Poisson structure $\Pi_{G\times G} = \Pi \oplus \Pi$ on $G \times G$.
\end{definition}

It is well-known~\cite{Drinfeld1983hamiltonian} that Poisson--Lie (PL) structures on a (connected and simply connected) Lie group $G$ are in one-to-one correspondence with \emph{Lie bialgebra structures $(\mathfrak g, [\cdot,\cdot], \delta)$}, where the skewsymmetric cocommutator map $
\delta:{\frak g}\to \bigwedge^2 \mathfrak{g}
$
fulfills the two following conditions:
\begin{itemize}
\item (i) $\delta$ is a 1-cocycle with values in $\bigwedge^2 \mathfrak{g}$, {\em  i.e.},
\be
\delta([X,Y])=[\delta(X),\,  Y\otimes 1+ 1\otimes Y] + 
[ X\otimes 1+1\otimes X,\, \delta(Y)] ,\qquad \forall \,X,Y\in
\frak g.
\label{1cocycle}
\ee
\item (ii) The transpose map $^t \delta = [\cdot,\cdot]_*:{\frak g}^\ast\wedge {\frak g}^\ast \to {\frak g}^\ast$ defines a Lie bracket on $\frak g^\ast$.
\end{itemize}
Note that the notion of Lie bialgebra is self-dual: for any Lie bialgebra $(\frak g, [\cdot,\cdot] , \delta)$ there exists a dual Lie bialgebra $(\frak g^\ast, [\cdot,\cdot]_*, \delta^*)$ where the Lie algebra structure on $\frak g^\ast$ is defined by $^t \delta = [\cdot,\cdot]_*$, and the dual cocommutator map $\delta^*$ is given by dualizing the Lie algebra relations in $\frak g$, i.e. $\delta^* = ^t [\cdot,\cdot]$.

The PL group-Lie bialgebra one-to-one correspondence is analogous to the one between Lie groups $G$ and Lie algebras $\mathfrak g$. Thus, the cocommutator $\delta$ is obtained essentially by differentiating the Poisson bivector $\Pi$ at the identity element $e \in G$. In order to see that, let us consider the map
\begin{equation}
\begin{split}
\label{etapoise}
\eta : G &\rightarrow \bigwedge^2 \mathfrak{g} \\
g &\rightarrow \left( (R_{g^{-1}})_{*} \otimes (R_{g^{-1}})_{*} \right) \Pi(g),
\end{split}
\end{equation}
which is just the right translation of the Poisson bivector $\Pi$ to the identity $e \in G$. The fact that $\Pi$ is compatible with the group multiplication implies that $\eta$ is in fact a cocycle of $G$ with values in $\bigwedge^2 \mathfrak{g}$, i.e.
\begin{equation}
\label{eq:cocycle_eta}
\eta (g_1 g_2) = \eta (g_1) + \mathrm{Ad}_{g_1} \eta (g_2),
\end{equation}
for all $g_1,g_2 \in G$. The derivative at the identity $e \in G$ of $\eta$ defines the cocommutator map 
\begin{equation}
\begin{split}
\delta :  \mathfrak{g} &\rightarrow \bigwedge^2 \mathfrak{g} \\
X &\rightarrow \frac{d}{dt} \bigg |_{t=0} \eta (e^{t X}) =  \frac{d}{dt} \bigg |_{t=0} \left( (R_{e^{-t X}})_{*} \otimes (R_{e^{-t X}})_{*} \right) \Pi(e^{t X}).
\end{split}
\end{equation}
Thus the compatibility of $\Pi$ with the group multiplication, or equivalently equation \eqref{eq:cocycle_eta}, induces the 1-cocyle condition (i) for the cocommutator $\delta$. Moreover, the co-Jacobi condition (ii) is simply the tangent remnant of the Jacobi condition for $\Pi$.

In the rest of this paper, a prominent role will be played by Poisson-Lie subgroups, so let us give a precise definition. 
\begin{definition}\cite{ChariPressley1994}
\label{def:PLsubgroup}
A \emph{Poisson-Lie subgroup} $(H,\Pi_H)$ of a Poisson-Lie group $(G,\Pi_G)$ is a Lie subgroup which is a Poisson submanifold of $(G,\Pi_G)$, i.e. the canonical inclusion $i : H \hookrightarrow G$ is a Poisson map and $(\Pi_G)_{h} \in \bigwedge^2(T_h H)$ for all $h \in H$.
\end{definition} 
In particular, it is clear that $(H,\Pi_H)$ is a PL group itself and thus, at the tangent level, we can write $\delta (\mathfrak h) \subseteq \mathfrak h \wedge \mathfrak h$, where $\mathfrak h = \mathrm{Lie} (H)$. However, this definition is quite restrictive and the following notion will become more relevant for our purposes.

\begin{definition} \cite{Lu1990thesis}
\label{def:coisotropicPLsubgroup}
Let $(G,\Pi_G)$ be the unique connected and simply connected PL group with Lie bialgebra $(\mathfrak g, [\cdot,\cdot], \delta)$ and let $H$ be a Lie subgroup of $G$ with Lie algebra $\mathfrak h = \mathrm{Lie} (H)$. We say that $(H,\Pi_H)$ is a \emph{coisotropic Poisson-Lie subgroup} of $(G,\Pi_G)$ if $\mathfrak h$ is a left coideal of $\mathfrak g$, i.e. $\delta (\mathfrak h) \subseteq \mathfrak h \wedge \mathfrak g$.
\end{definition}
Indeed, PL subgroups are particular cases of coisotropic PL subgroups.

A description in terms of coordinates will be useful in what follows. Thus, let $(G,\Pi_G)$ be a PL group and $(H,\Pi_H)$ a coisotropic PL subgroup. Then we can always parametrize locally $G$ by introducing a set of local coordinates $\{x,\xi\}$ in such a way that $\{\xi\}$ are local coordinates on $H$. In terms of these local coordinates, the PL bracket on $G$ will be written generically in the form
\be
\{ x, x\}_\Pi= A(x,\xi)
\qquad
\{ x, \xi\}_\Pi= B(x,\xi)
\qquad
\{ \xi,\xi\}_\Pi= C(x,\xi),
\label{plgroup}
\ee
where $A,B$ and $C$ belong to $\mathcal C^\infty (G)$, and the linearization (in terms of the local coordinates $\{\xi,x\}$) of these $\Pi$ brackets will be just the Poisson bracket analogues of the dual Lie algebra defined by $^t \delta = [\cdot,\cdot]_*$. 

Finally, we recall that PL groups $(G,\Pi)$ are the `semiclassical' counterparts of quantum groups (in other words, quantum groups are quantizations of PL groups as Hopf algebras, for a detailed account see~\cite{Drinfeld1987icm,ChariPressley1994,Majid1995book}). Therefore, the fundamental relations defining the quantum group $G_q$ will be obtained as the quantization of the $\Pi$ algebra~\eqref{plgroup}, and the Lie algebra $(\frak g^\ast, [\cdot,\cdot]_*)$ will just be the first order of such noncommutative algebra $G_q$. In this paper we will denote by the symbols $\{\hat x,\hat \xi\}$ to the noncommutative analogues of $\{x,\xi\}$, and we will refer to them as the `quantum coordinates'.

\section{Homogeneous spaces and Poisson homogeneous spaces}

The aim of this section is twofold: firstly, we recall some basic notions regarding the geometry of homogeneous spaces. Secondly, we present the basic concepts and definitions concerning Poisson homogeneous spaces that we will be using throughout the rest of the work. In the next section, we will show that these two independent topics are indeed related by a generalization of the `quantum duality principle'~\cite{Drinfeld1987icm, Semenov-Tyan-Shanskii1992, Gavarini2002, CG2006qd}. 

\subsection{Homogeneous spaces}

The geometric notion of a space for which each point is equivalent to any other point is encoded in the definition of a homogeneous space:

\begin{definition}
\label{def:HS}
A \emph{homogeneous space} for the Lie group $G$ is a smooth manifold $M$ endowed with a smooth and transitive action of $G$. 
\end{definition}
It is well-known that cosets of the form $G/H_m$ are models for homogeneous spaces for the Lie group $G$, where $m \in M$ is a point of the homogeneous space $M$ and $H_m$ is the stabilizer of $m$ in $G$. Moreover, isotropy subgroups for different points $m,m' \in M$ are conjugate, and thus isomorphic. This is the mathematical description of the fact that in a homogeneous space all points are equivalent. Nevertheless, when we take a concrete coset we are indeed  fixing the origin of the resulting space. As it will be relevant when introducing Poisson structures, we introduce the following

\begin{definition}
\label{def:pointedHS}
A \emph{pointed homogeneous space} is a coset $G/H$, where $H$ is a (fixed) closed subgroup of $G$.
\end{definition}

For a pointed homogeneous space $G/H$ we will denote its origin by $eH$, $e$ being the identity element of the Lie group $G$. We remark that this is the only point whose isotropy subgroup is $H$. 

The main property of homogeneous spaces is that $G$ acts smoothly and transitively on them, and thus $G$-invariant structures should be naturally considered. Given that in general this is quite a complicated task, we usually look at those spaces that satisfy some extra conditions, namely the so-called \emph{reductive} and \emph{symmetric} homogeneous spaces.

\subsection{Reductive and symmetric homogeneous spaces}

\begin{definition}
\label{def:reductivespace}
For $M=G/H$ a homogeneous space, we call $M$ \emph{reductive} if $\mathfrak{g} = \mathrm{Lie}\, (G)$ (as a vector space) can be decomposed into a direct sum of $\mathfrak{h} = \mathrm{Lie}\, (H)$ and an $\mathrm{Ad}_H$-invariant subspace $\mathfrak{t}$, that is, if
\begin{itemize}
\item[i) ] $\mathfrak{g} = \mathfrak{h} \oplus \mathfrak{t}, \qquad \mathfrak{h} \cap \mathfrak{t} = 0$,
\item[ii) ] $\mathrm{Ad}_H \mathfrak{t} \subseteq \mathfrak{t}.$
\end{itemize}
Condition $ii)$ implies that $\mathfrak{t}$ is $\mathrm{ad}_\mathfrak{h}$-invariant, namely
\begin{itemize}
\item[ii)' ] $[\mathfrak{h},\mathfrak{t}] \subseteq \mathfrak{t}$.
\end{itemize}
If $H$ is connected, then $ii)'$ implies $ii)$. From now on, this will be always the case in this paper. Thus, for reductive homogeneous spaces, the structure of the Lie algebra $\mathfrak g$ can be written as
\begin{equation}
\label{eq:reducLiealg}
[\mathfrak h,\mathfrak h] \subseteq \mathfrak h, \qquad \quad [\mathfrak h,\mathfrak t] \subseteq \mathfrak t, \qquad \quad [\mathfrak t,\mathfrak t] \subseteq \mathfrak h \oplus \mathfrak t.
\end{equation}
\end{definition}
The simplicity of reductive homogeneous spaces can be traced back to the fact that $\mathrm{Ad}_H$-invariance for $\mathfrak{t}$ implies that we can further identify $T_o (M) =T_{eH}(G/H) \simeq \mathfrak{g}/\mathfrak{h} \simeq \mathfrak{t}$. 

\paragraph{Remark.}
We say that a Lie algebra of the form \eqref{eq:reducLiealg} is $(\mathfrak{h}, \mathfrak t)$-reductive.

\begin{definition}
\label{def:symmetric}
A \emph{symmetric space} is a triple $(G,H,\sigma)$ consisting of a connected Lie group, a closed subgroup $H$ of $G$ and an involutive automorphism $\sigma$ of $G$ such that $H$ lies between $G_\sigma$ and the identity component of $G_\sigma$.
\end{definition}
Obviously, if $G_\sigma$ is connected then $H=G_\sigma$. The tangent space of a symmetric space naturally inherits the structure of a symmetric Lie algebra.
\begin{definition}
\label{def:symliealg}
A \emph{symmetric Lie algebra} is a triple $(\mathfrak g,\mathfrak h,\sigma)$ consisting of a Lie algebra $\mathfrak g$, a Lie subalgebra $\mathfrak h$ and an involutive automorphism $\sigma$ of $\mathfrak g$, where $\mathfrak h=\{ X \in \mathfrak g \, | \, \sigma (X)=X \}$.\end{definition}
For any symmetric Lie algebra, involutivity of $\sigma$ directly implies that, as a linear map, its only possible eigenvalues are $\pm 1$. If we call $\mathfrak h$ to the eigenspace corresponding to $+1$ and $\mathfrak t$ to the eigenspace corresponding to $-1$, i.e. 
\begin{equation}
\begin{split}
\sigma (X) &= X, \qquad \quad \forall X \in \mathfrak h \\
\sigma (X) &= -X, \qquad \;  \forall X \in \mathfrak t \\
\end{split}
\end{equation}
then it is straightforward to prove that
\begin{equation}
\label{eq:symmLiealg}
[\mathfrak h,\mathfrak h] \subseteq \mathfrak h, \qquad\quad [\mathfrak h,\mathfrak t] \subseteq \mathfrak t, \qquad\quad [\mathfrak t,\mathfrak t] \subseteq \mathfrak h.
\end{equation}
\begin{proposition}\cite{KobayashiNomizu1969-2}
\label{prop:AdHinvsplit}
Let $(G,H,\sigma)$ be a symmetric space with tangent symmetric Lie algebra $(\mathfrak g,\mathfrak h,\sigma)$. Then the splitting $\mathfrak g = \mathfrak h \oplus \mathfrak t$ is $\mathrm{Ad}_H$-invariant, i.e. $\mathrm{Ad}_H \mathfrak t \subset \mathfrak t$.
\end{proposition}
The Proposition above, together with \eqref{eq:symmLiealg}, shows that every symmetric Lie algebra is $(\mathfrak{h}, \mathfrak t)$-reductive with $\mathfrak{h}$ and $\mathfrak t$ being the $+1$ and $-1$ eigenspaces for $\sigma$, respectively. Note that for every symmetric algebra $\mathfrak h \cap \mathfrak t = 0$ trivially. In the same way, \emph{every symmetric homogeneous space is reductive}.

Summarizing, each one of the three conditions listed above (isotropy with respect to $H$, reductivity and symmetry) removes  one specific type of contribution within the most generic Lie algebra brackets for $\frak g$, namely: 
\begin{equation}\label{Symmetric_space_conditions}
[\frak h,\frak h] \subseteq  \frak h \oplus \bcancel{\frak t }_\text{~isotropy} 
\,, \qquad
[\frak h,\frak t] \subseteq  \bcancel{\frak h}_\text{~reductive space} \oplus \frak t
\,, \qquad
[\frak t,\frak t] \subseteq  \frak h \oplus \bcancel{\frak t}_\text{~symmetric space}
\, .
\end{equation}

\subsection{Poisson homogeneous spaces}

The notion of a homogeneous space can be extended to the category of Poisson manifolds. The key idea is to add, in a compatible way, a Poisson structure to both the Lie group (becoming a PL group) and to the homogeneous space. 

\begin{definition}
For any Poisson-Lie group $(G,\Pi)$, a \emph{$G$-Poisson homogeneous space ($G$-PHS)} is a homogeneous space for $G$ endowed with a Poisson structure $\pi$ such that the action of $G$ on $M$ is a Poisson map with respect to the PL group structure $\Pi$ on $G$ and the product Poisson structure $\Pi \oplus \pi$ on $G \times M$. When no confusion concerning $G$ is possible, we will use PHS as an abbreviation.
\end{definition}

The compatibility between the Poisson structures on the Lie group $G$ and on the homogeneous manifold $M$  is called the \emph{covariance condition}. However, an important feature arises when a Poisson structure is defined onto $M$: despite this space is homogeneous from a geometric perspective (every point is \emph{equivalent}), the Poisson structure is by no means homogeneous in the sense that the Poisson bracket depends on the point of the manifold that we are considering (and the same will happen when the corresponding quantum homogenous space is constructed). 

Most of the literature concerning PHS has focused in the problem of classifying the different PHS for a given PL group. In this respect, in a seminal paper \cite{Drinfeld1993}, Drinfel'd proved that PHS are in one-to-one correspondence with orbits of a natural action of the Lie group $G$ on the variety of Lagrangian subalgebras $\mathcal L(D(G))$ of the Drinfel'd double $D(G)$. Explicit examples of PHS  as well as some classification results can be found, for instance, in \cite{Ciccoli1996qplanes,Karolinsky1995,Karolinsky1996,Karolinsky1997,Karolinsky1998,Karolinsky1998a,Lu2000,KS2002dynamical,Karolinsky2004,Karolinsky2005,Zakrzewsiki1995phs}. It is worth stressing the existence of a relevant relation between PHS and the Dynamical Classical Yang-Baxter equation (DCYBE), which allowed the quantization of certain PHS in terms of $G$-equivariant star products (see~\cite{Karolinsky2005}) and has been used, for instance, in the context of gauge fixing in $(2+1)$-gravity \cite{MS2011gaugefixing,MS2014dynamical}.

We stress that the viewpoint adopted in this work is fundamentally different, since we do not face the generic classification problem for PHS, and we focus on the analysis of Poisson structures defined on certain (and fixed) PHS. To this aim, we introduce the following

\begin{definition}
Let $(G,\Pi)$ be a Poisson-Lie group and $(H,\Pi_H)$ a \emph{coisotropic Poisson-Lie subgroup} of $(G,\Pi)$. A \emph{$H$-coisotropic pointed Poisson homogeneous space} is a Poisson manifold $(M,\pi)$ diffeomorphic to $G/H$ such that the canonical projection $p:G \to G/H$ is a Poisson map.
\end{definition}

In the above definition, the Lie subgroup $H$ is the stabilizer of the origin $eH$ of M. If we consider a different point $m \in M$, then the coset $G/H_m$ would also be diffeomorphic to $M$ and $H_m$ would be conjugate to $H$. However, the Poisson structure could be, in general, different (i.e. associated to a different orbit of $G$ on $\mathcal L(D(G))$), although this feature will not be considered in what follows. 

Note that coisotropic pointed PHS, also called Poisson quotients \cite{Zakrzewsiki1995phs,CG2006qd}, have the remarkable property that they are never symplectic, since $\pi_{eH} = 0$. Among coisotropic pointed PHS, those for which $(H,\Pi_H)$ is a Poisson-Lie subgroup of $(G,\Pi)$ (see Definition \ref{def:PLsubgroup}) are particularly simple.

We also stress that PHS are the semiclassical counterparts of noncommutative quantum homogeneous spacetimes $M_q$, which are covariant under quantum group $G_q$ (co)actions. In fact, the introduction of the bracket $\pi$ on a given homogeneous spacetime can be interpreted as the most natural way to construct a noncommutative spacetime, after a suitable quantization procedure is performed. Moreover, such quantization for $(M,\pi)$ will give rise to the quantum homogeneous space $M_q$, which will be invariant under the appropriate (co)action of the quantum group $G_q$. Concerning the theory of quantum homogeneous spaces, we refer the reader to~\cite{DijkhuizenKoornwinder1994,Ciccoli1996qplanes,Karolinsky2005} and references therein.


\section{Complementary dual Poisson homogeneous spaces}

Given a PL group $(G,\Pi)$ and a PL subgroup $(H,\Pi_H)$, it directly follows from Definition \ref{def:coisotropicPLsubgroup} that the coset $G/H$ with the inherited Poisson structure is a pointed coisotropic PHS if and only if 
\begin{equation}
\label{eq:coisotropycond}
\delta (\mathfrak h) \subseteq \mathfrak h \wedge \mathfrak g . 
\end{equation}

\begin{definition}
Let $(\mathfrak g, [\cdot,\cdot], \delta)$ be a Lie bialgebra, $\mathfrak h \subseteq \mathfrak g$ a Lie subalgebra and $\mathfrak t \subseteq \mathfrak g$ its orthogonal complement. We say that the Lie bialgebra $(\mathfrak g, [\cdot,\cdot], \delta)$ is $(\mathfrak h,\mathfrak t)$-coisotropic if the condition~\eqref{eq:coisotropycond} is satisfied. If there is no possible confusion concerning $\mathfrak h$ and $\mathfrak t$, we simply say that the Lie bialgebra $(\mathfrak g, [\cdot,\cdot], \delta)$ is coisotropic.
\end{definition}

It is important to stress that this condition is not invariant under conjugation, which means that by considering $G/H'$, where $H'$ is the stabilizer of a different point and thus a conjugate subgroup to $H$, the resulting pointed PHS could be non coisotropic. However, as we have remarked before, in the present work this possibility will not be considered since the isotropy subgroup of the homogeneous space (and therefore its origin), will be completely fixed. 

In~\cite{Lu1990thesis} it is proven that the coisotropy condition is equivalent to the fact that the \emph{annihilator of $\mathfrak h$ in $\mathfrak g^\ast$}, namely
\begin{equation}
\mathfrak h^\perp = \{ \phi \in \mathfrak g^* \, | \, \phi (X)=0, \; \forall X \in \mathfrak h \} ,
\end{equation}
 is a Lie subalgebra of $\mathfrak g^*$.  This implies that the following notion can be introduced:
 
\begin{definition} \cite{CG2006qd} For every pointed $H$-coisotropic PHS, the {\em complementary dual space} $M^\perp=G^*/H^\perp$ is defined  as the coset space of the PL group $G^\ast$ with respect to the isotropy subgroup $H^\perp$ whose Lie algebra is the annihilator $\mathfrak h^\perp$. 
\end{definition}

By construction, $M^\perp$ is a homogeneous space for $G^*$. Moreover $M^\perp$ is a $H^\perp$-coisotropic pointed PHS, because $\mathfrak h$ is a Lie subalgebra of $\mathfrak g$,  and this subalgebra property is translated in terms of the dual cocommutator as
\begin{equation}
\delta^* (\mathfrak h^\perp) \subseteq \mathfrak h^\perp \wedge \mathfrak g^* ,
\end{equation}
which is just the coisotropy condition for $\mathfrak h^\perp$ when embedded in $ \mathfrak g^*$. This straightforward result encodes relevant conceptual information since the original Poisson structure on $M$ turns out to be related with the geometry of the associated complementary dual homogeneous space. Obviously, the dimensions of $M$ and $M^\perp$ are connected through the relation
\begin{equation}
\mathrm{dim} \, M + \mathrm{dim} \, M^\perp = \mathrm{dim} \, G = \mathrm{dim} \, G^* .
\end{equation}
Thus there exists a deep correspondence (mediated by the Lie bialgebra structure) between two different Poisson homogeneous spaces $M$ and $M^\perp$, which have in general different geometric properties and dimensionality.

\subsection{Local coordinates}

The coordinates $\{x,\xi\}$ on $G$ are such that the coordinate functions $\{\xi\}$ represent local coordinates on the isotropy subgroup $H$. On the other hand,  the coordinates on $M=G/H$ can be identified with the local coordinates $x$ which are dual to the translation generators $T_i$ spanning $\mathfrak t$, provided that the former are well defined functions in the quotient space, i.e. $x \in \mathcal C^\infty (M)$, which is tantamount to say that $x$ are invariant under either the right or left action (depending on whether we consider left or right cosets) of $G$ onto itself.

After having at hand such a set of appropriate coordinates $\{x,\xi\}$ on $G$, the problem of finding a homogeneous Poisson structure $\pi$ on $M$ that is compatible with the group action $\rhd: G\times M\to M$ can be solved by composing the  PL structure on $G$ given by~\eqref{plgroup} with the canonical projection $p:G \rightarrow G/H$, $g \rightarrow gH$ provided that $H$ is a coisotropic PL subgroup. This guarantees that the $x$ coordinates do generate  a Poisson subalgebra within $\Pi$, which means that the function $A$ from~\eqref{plgroup} depends only on $x$, and the PHS bracket reads
\be
\{ x, x\}_\pi\equiv \{ x, x\}_\Pi=A(x).
\label{piproy}
\ee
As a consequence, the linearization of $\Pi$~\eqref{piproy} has to give rise to the Lie subalgebra on $\mathfrak g^\ast$ given by
\be
[\frak h^\perp,\frak h^\perp]_* \subseteq  \frak h^\perp,
\label{coisotropy}
\ee
and this dual condition can be translated into the Lie bialgebra cocommutator as the non-existence of $\frak t\wedge \frak t$ terms in $\delta(\frak h)$, which is thus constrained to be of the form
\be
\delta(\frak h) \subseteq \frak h \wedge \frak h \oplus \frak h \wedge \frak t \, ,
\label{coisotropyext}
\ee
which is just the coisotropy condition~\eqref{eq:coisotropycond}.

In the following we will be interested in describing the complementary dual spaces $M^\perp$  in terms of local coordinates. In order to do that, we firstly introduce a basis $\{\hat\xi , \hat x \}$  for the dual Lie algebra $\frak g^\ast$ given by $\frak t^\perp =\mbox{span}\{\h^i\}$ and $\frak h^\perp =\mbox{span}\{\t^j\}$
together with the following pairing with the generators of $\frak g$, where $\frak h=\mbox{span}\{H_i\}$ and $\frak t=\mbox{span}\{T_j \}$:
\begin{equation}\label{Duality_relations}
\langle \h^i , H_j \rangle = \delta^i_j \,, \qquad 
\langle \h^i ,T_j \rangle = 0 \,, \qquad
\langle \t^i , H_j \rangle = 0 \,, \qquad
\langle \t^i , T_j \rangle = \delta^i_j  \,.
\end{equation}
In order to have a set of local coordinate functions on $M^\perp=G^*/H^\perp$ we firstly introduce local coordinates $\{\xi^\ast,x^\ast\}$ on $G^*$ by exponentiating the coordinates on the Lie algebra $\frak g^\ast$ (thus defining the so-called \emph{exponential coordinates of the second kind} with respect to the Lie algebra basis $\{\hat\xi,\hat x\}$). Then, the coset structure allows us to define local coordinates on $M^\perp$ by composing the functions $\xi^\ast$ with the canonical projection $p^\perp : G^* \to M^\perp$.

It is clear that the coisotropy condition stated as~\eqref{coisotropy} is just a dual counterpart of the isotropy condition for the subalgebra $\frak h$ and thus the basis elements $\hat x^j$ dual to the translations $T_i$ close a Lie subalgebra~\eqref{coisotropy} within the dual Lie bracket $[\cdot,\cdot ]_*$ on $\frak g^\ast$. As we will see in Section 5, further coreductivity and cosymmetry conditions can be introduced as refinements of the notion of coisotropic Lie bialgebras, and they will arise when the dual Lie algebra $(\frak g^\ast, [\cdot,\cdot ]_*)$ is constrained to match with the reductivity and symmetry conditions for $M^\perp$ as a $G^\ast$-homogeneous space.


\section{Coreductivity and cosymmetry}


In the rest of this paper we investigate the consequences for the Poisson structure on $M$ that are derived by imposing that the `complementary dual' $M^\perp$ is either a reductive or a symmetric homogeneous space. As we will see in the following, such conditions impose severe restrictions on the cocommutator $\delta$ of the Lie bialgebra associated to the PL structure $\Pi$ on $G$ that can be considered.

For any pointed homogeneous space $G/H$, we can always write $\frak g = \frak h \oplus \frak t$ (recall that $\mathfrak h$ is fixed from the beginning). In the case of $G/H$ being a reductive or symmetric homogeneous space, this splitting is $Ad_H$-invariant. Therefore, the generic form for the cocommutator $\delta$ associated to a given PL structure on $G$ is written as
\begin{equation}
\begin{gathered}
\delta(\frak h) \subseteq \frak h \wedge \frak h \oplus \frak h \wedge \frak t \oplus \frak t \wedge \frak t \,,
\\
\delta(\frak t) \subseteq \frak h \wedge \frak h \oplus \frak h \wedge \frak t \oplus \frak t \wedge \frak t \, .
\end{gathered}
\label{coco}
\end{equation}
At this point the coisotropy condition \eqref{eq:coisotropycond} can be expressed as the cancellation of the term $\frak t \wedge \frak t$ in $\delta(\frak h)$, i.e~\eqref{coisotropyext}, 
and in terms of the dual Lie algebra, this condition implies that $\frak h^\perp$ is a Lie subalgebra of $\mathfrak g^*$. Therefore,~\eqref{coisotropy} holds
and thus the complementary dual space $M^\perp =G^*/H^\perp$ is well defined, since $H^\perp$ is well-defined as its isotropy subgroup. Moreover, $M^\perp$ is by construction a Poisson homogeneous space, whose Poisson bracket $\pi^\ast$ on $\mathcal C^\infty(M^\perp)$ will be given by the canonical projection of the dual PL structure $\Pi^\ast$ onto the $\xi^\ast$ coordinates parametrizing $M^\perp$. 

Recall that $\Pi^\ast$ has as its  linearization the dual Lie bialgebra structure $(\frak g^\ast,  [\cdot,\cdot]_*, \delta^*)$ whose cocommutator map comes from the commutation rules of the Lie algebra $\frak g$, hence in particular the PHS defined by $(M^\perp,\pi^\ast)$ is always $(\frak h^\perp,\frak t^\perp)$-coisotropic, since
\begin{equation}
\delta^* (\frak h^\perp) \subseteq  \frak h^\perp \wedge \frak g^\ast \, ,
\end{equation}
and this is due to the fact that  $[\frak h,\frak h] \subseteq \frak h$. Therefore, $(\mathfrak h,\mathfrak t)$-coisotropy for $M$ is equivalent to the fact that $\frak h^\perp$ is a Lie subalgebra, and $(\mathfrak h^\perp,\mathfrak t^\perp)$-coisotropy for $M^\perp$ arises since $\frak h$ is a Lie subalgebra. Mofeover, we stress that the full construction is self-dual, and the complementary dual PHS of $M^\perp$ with respect to the Lie bialgebra $(\frak g^\ast,  [\cdot,\cdot]_*, \delta^*)$ will be just $(M^\perp)^\perp = M$.

\subsection{Complementary dual reductive PHS}

If we now impose that the complementary dual space  $M^\perp$ to $M$ is not only a homogeneous space but a reductive space, condition ii) from Definition \eqref{def:reductivespace} implies that
\begin{equation}
[\frak h^\perp,\frak t^\perp] \subseteq \frak t^\perp ,
\end{equation} 
where
\begin{equation}
\mathfrak t^\perp = \{ \phi \in \mathfrak g^* \, | \, \phi (X)=0, \; \forall X \in \mathfrak t \} .
\end{equation}
This is translated in the original Lie bialgebra structure by
 imposing that no $\frak h \wedge \frak t$ term can be contained within $\delta(\frak t)$, namely
\begin{equation}\label{coredt}
\delta(\frak t) \subseteq \frak h \wedge \frak h \oplus  \frak t \wedge \frak t \,.
\end{equation}
By analogy to the coisotropy condition, we call~\eqref{coredt} the \emph{coreductivity condition}. 

\begin{definition}
Let $(\mathfrak g, [\cdot,\cdot], \delta)$ be a Lie bialgebra, $\mathfrak h \subseteq \mathfrak g$ a Lie subalgebra and $\mathfrak t \subseteq \mathfrak g$ its orthogonal complement. We say that the Lie bialgebra $\mathfrak g$ is $(\mathfrak h,\mathfrak t)$-coreductive if the condition~\eqref{coredt} is satisfied. If there is no possible confusion concerning $\mathfrak h$ and $\mathfrak t$, we simply say that the Lie bialgebra $(\mathfrak g, [\cdot,\cdot], \delta)$ is coreductive.
\end{definition}

If such condition is imposed, the structural relations for the dual Lie algebra $(\frak g^\ast,[\cdot,\cdot]_*)$ read
\begin{equation}\label{Symmetric_space_conditions_2}
[\frak h^\perp,\frak h^\perp]_* \subseteq \frak h^\perp 
\,, \qquad
[\frak h^\perp,\frak t^\perp]_* \subseteq  \frak t^\perp
\,, \qquad
[\frak t^\perp,\frak t^\perp]_* \subseteq \frak h^\perp \oplus \frak t^\perp
\,,
\end{equation} 
since we want $\frak g^\ast$ to be the Lie algebra of a reductive homogeneous space for $G^*$ with isotropy subgroup given by $H^\perp$. As we will see in Section 7, the coreductivity condition will give rise to a strong constraint for the type of uncertainty relations that can be associated, after quantization, to the noncommutative coordinates of $M_q$.


\subsection{Complementary dual symmetric PHS}

A further condition can be considered if we impose the complementary dual space $M^\perp$ to be a symmetric space. This implies the existence of an involutive automorphism $\sigma^\ast$ leaving invariant the generators of the isotropy subgroup $\frak h^\perp$, namely
\begin{equation}
\sigma^\ast(\frak h^\perp) = \frak h^\perp \,, \qquad \sigma^\ast(\frak t^\perp) = - \frak t^\perp \,.
\end{equation}
This implies that the Lie brackets for $\frak g^\ast$ have to be of the form
\begin{equation}
[\frak h^\perp,\frak h^\perp]_* \subseteq \frak h^\perp 
\,, \qquad
[\frak h^\perp,\frak t^\perp]_* \subseteq  \frak t^\perp
\,, \qquad
[\frak t^\perp,\frak t^\perp]_* \subseteq \frak h^\perp 
\, .
\label{dualtodo}
\end{equation} 
This constraint, rewritten in terms of the original Lie bialgebra $(\mathfrak g, [\cdot,\cdot], \delta)$ leads to the following simplified cocommutator
\begin{equation}
\delta(\frak h) \subseteq  \frak h \wedge \frak t\,, \qquad\qquad 
\delta(\frak t) \subseteq \frak h \wedge \frak h \oplus  \frak t \wedge \frak t \,.
\end{equation}
In this way, the new constraint for the Lie bialgebra $(\mathfrak g, [\cdot,\cdot], \delta)$ resulting from $M^\perp =G^*/H^\perp$ being a symmetric space, in addition to a reductive space, is given by the \emph{cosymmetry condition}
\begin{equation}
\label{cosymm}
\delta(\frak h) \subseteq  \frak h \wedge \frak t\,.
\end{equation}

\begin{definition}
Let $(\mathfrak g, [\cdot,\cdot], \delta)$ be a Lie bialgebra, $\mathfrak h \subseteq \mathfrak g$ a Lie subalgebra and $\mathfrak t \subseteq \mathfrak g$ its orthogonal complement. We say that the Lie bialgebra $\mathfrak g$ is $(\mathfrak h,\mathfrak t)$-cosymmetric if the cosymmetry condition~\eqref{cosymm} is satisfied. If there is no possible confusion concerning $\mathfrak h$ and $\mathfrak t$, we simply say that the Lie bialgebra $(\mathfrak g, [\cdot,\cdot], \delta)$ is cosymmetric.
\end{definition}

All the previous results can be summarized in terms of the algebraic properties of the initial Lie bialgebra $(\mathfrak g, [\cdot,\cdot], \delta)$ as follows:

\begin{itemize}
\item An $(\mathfrak h,\mathfrak t)$-coisotropic Lie bialgebra  associated to a generic pointed coisotropic PHS $M=G/H$ is of the form:
\begin{equation}
\begin{cases}
[\frak h,\frak h] \subseteq  \frak h  
\,, \qquad
[\frak h,\frak t] \subseteq \frak h \oplus \frak t
\,, \qquad
[\frak t,\frak t] \subseteq \frak h \oplus \frak t
\, , \\
\delta(\frak h) \subseteq \frak h \wedge \frak h \oplus \frak h \wedge \frak t  \,, \qquad\qquad \delta(\frak t) \subseteq  \frak h \wedge \frak h \oplus \frak h \wedge \frak t \oplus \frak t \wedge \frak t \,.
\end{cases}
\end{equation}

\item An $(\mathfrak h,\mathfrak t)$-coreductive Lie bialgebra  associated to a reductive $M$ is of the form:
\begin{equation}
\begin{cases}
[\frak h,\frak h] \subseteq \frak h  
\,, \qquad
[\frak h,\frak t] \subseteq \frak t
\,, \qquad
[\frak t,\frak t] \subseteq \frak h \oplus \frak t
\, , \\
\delta(\frak h) \subseteq \frak h \wedge \frak h \oplus \frak h \wedge \frak t  \,, \qquad\qquad \delta(\frak t) \subseteq  \frak h \wedge \frak h \oplus \frak t \wedge \frak t \,.
\end{cases}
\end{equation}

\item Finally, an $(\mathfrak h,\mathfrak t)$-cosymmetric Lie bialgebra  associated to a symmetric $M$ would be:
\begin{equation}
\begin{cases}
[\frak h,\frak h] \subseteq  \frak h  
\,, \qquad
[\frak h,\frak t] \subseteq  \frak t
\,, \qquad
[\frak t,\frak t] \subseteq  \frak h 
\, . \\
\delta(\frak h) \subseteq  \frak h \wedge \frak t  \,, \qquad\qquad \delta(\frak t) \subseteq  \frak h \wedge \frak h \oplus \frak t \wedge \frak t \,.
\end{cases}
\end{equation}

\end{itemize}

We stress that this algebraic characterization for each kind of PHS is self-dual in the sense that the complementary dual PHS $M^\perp = G^*/H^\perp$ satisfies the same kind of relations, provided we perform the following replacements 
\begin{equation}
\mathfrak h \to \mathfrak h^\perp \,, \qquad
\mathfrak t \to \mathfrak t^\perp \, , \qquad 
[\cdot,\cdot] \to [\cdot,\cdot]_* \, , \qquad 
\delta \to \delta^* .
\end{equation}
However, both the geometry and the Poisson structure of $M$ and $M^\perp$ are in general very different ones. Nevertheless, the conditions arising from being of coreductive and cosymmetric pointed PHS imply that certain important properties of the geometry of both the original and the `complementary dual' homogeneous space can be preserved. For instance, starting with a reductive PHS and considering a coreductive Lie bialgebra, we guarantee that the study of $K$-invariant structures on both $M$ and $M^\perp$ is far simpler than the general case, as we will see in section 7. Similarly, if $M$ is a symmetric space and we consider a cosymmetric Lie bialgebra, we guarantee that both $M$ and $M^\perp$ are endowed with a point symmetry which is globally defined. Moreover, these two notions illustrate in a really simple way how the Poisson structure of a given PHS is linked with the geometry of a different (but related) space. 

A further motivation in order to study these two notions is the fact that some of the most interesting homogeneous spaces are indeed reductive and symmetric ones. This is the case for example for maximally symmetric Lorentzian homogeneous spaces, whose Poisson homogeneous analogues and their quantizations are typically studied in applications to quantum gravity. Therefore, in the following section all these notions will be exemplified by considering coreductive and cosymmetric structures for maximally symmetric Lorentzian homogeneous spaces in (2+1) and (3+1) dimensions.



\section{Coreductive Lorentzian Lie bialgebras}

In terms of the cosmological constant parameter $\Lambda$, the  three  maximally symmetric  (3+1)  Lorentzian
spacetimes are:
\begin{itemize}
\item $\Lambda<0$: Anti de Sitter spacetime ${\mathbf
{AdS}}^{3+1} \equiv  {\rm SO}(3,2)/{\rm  SO}(3,1)$.

\item $\Lambda>0$: de Sitter spacetime ${\mathbf
{dS}}^{3+1} \equiv   {\rm  SO}(4,1)/{\rm SO}(3,1)$.

\item $\Lambda=0$: Minkowski spacetime ${\mathbf
M}^{3+1} \equiv  {\rm  ISO}(3,1)/{\rm  SO}(3,1)$.
\end{itemize}

The ${\frak {so}}(3,2)$,   ${\frak {so}}(4,1)$ and  ${\frak {iso}}(3,1)$ Lie algebras can be simultaneously written in terms of the cosmological constant $\Lambda$ as the one-parameter family of Lie algebras $\frak g_\Lambda$ with Lie brackets
\be
\begin{array}{lll}
[J_a,J_b]=\epsilon_{abc}J_c ,& \quad [J_a,P_b]=\epsilon_{abc}P_c , &\quad
[J_a,K_b]=\epsilon_{abc}K_c , \\[2pt]
\displaystyle{
  [K_a,P_0]=P_a  } , &\quad\displaystyle{[K_a,P_b]=\delta_{ab} P_0} ,    &\quad\displaystyle{[K_a,K_b]=-\epsilon_{abc} J_c} , 
\\[2pt][P_0,P_a]=-\Lambda \, K_a , &\quad   [P_a,P_b]=\Lambda\, \epsilon_{abc}J_c , &\quad[P_0,J_a]=0  ,
\end{array}
\label{aa}
\ee
where
$\{P_0,P_a, K_a, J_a\}$ are the generators of time translation, space translations, boost transformations and rotations, respectively. Here $a,b,c=1,2,3$ and sum over repeated indices will be assumed. 

The 
decomposition of $\frak g_\Lambda$ (as a vector space) is given by
\begin{equation}
 \frak g_\Lambda ={\mathfrak{h}}  \oplus  {\mathfrak{t}} ,\qquad 
{\mathfrak{h}  }={\rm span}\{ \>K,\>J \}\simeq \mathfrak{so}(3,1) ,\qquad
{\mathfrak{t}  }={\rm span}\{  P_0,\>P \}  ,
\end{equation} 
where $\mathfrak{h} $ is the Lorentz subalgebra of the stabilizer of the origin of Minkowski spacetime. Therefore, (A)dS and Minkowski spacetimes, which we will denote as $M_\Lambda$, are symmetric (and thus reductive) homogeneous spaces with $\frak h$ being the Lorentz and $\frak t$ the translation subalgebra, and the commutation rules~\eqref{aa} can be schematically summarized in the form
\begin{equation}
[\frak h,\frak h] \subseteq  \frak h  \,, \qquad
[\frak h,\frak t] \subseteq    \frak t
\,, \qquad
[\frak t,\frak t] \subseteq  \Lambda\, \frak h 
\, .
\label{lorschem}
\end{equation} 
In the Minkowski ($\Lambda\to 0$) case we have $[\frak t,\frak t] =0$, which means that $\frak t$ generates a normal subgroup and we have the well-known semidirect product structure arising for the Poincar\'e algebra $\frak g_0$. Hereafter the Lorentz Lie subalgebra $\mathfrak h$ will be fixed (from a purely Lie algebraic point of view it should be considered as a embedding of $\mathfrak h$ into $\mathfrak g_\Lambda$), and therefore no confusion should arise when stating claims about the coisotropy, coreductivity and cosymmetry of a Lorentzian Lie bialgebra $(\mathfrak g_\Lambda,[\cdot,\cdot],\delta)$.

\subsection{Lorentzian Lie bialgebras}

It seems natural to investigate how the conditions of coisotropy, coreductivity and cosymmetry define a very specific subset within the family of all possible Lie bialgebra structures for Lorentzian Lie algebras $\frak g_\Lambda$ with commutation rules of the form~\eqref{lorschem} (see~\cite{Zakrzewski1997,Stachura1998,BHM2014tallinn,BLT2016unified,MS2018constraints} for classification approaches to Lorentzian Lie bialgebras).

It is well-known~\cite{Zakrzewski1997} that in (2+1) and (3+1) dimensions all (A)dS and Poincar\'e Lie bialgebras $(\mathfrak g_\Lambda,[\cdot,\cdot],\delta)$ are coboundary ones, which means that all of them can be obtained through $r$-matrices in the form:
\begin{equation}
\delta(X)=[1 \otimes X + X \otimes 1, r],
\qquad
\forall\, X\in \frak g_\Lambda
\label{cocomm}
\end{equation}
with $r \in \mathfrak g_\Lambda \otimes \mathfrak g_\Lambda$ being a skew-symmetric solution of the modified Classical Yang-Baxter Equation (mCYBE). 
Let us consider a generic $r$-matrix written in the schematic form
\begin{equation}
r \subseteq \a \, \frak h \wedge  \frak h \oplus \b \, \frak h \wedge  \frak t \oplus \c \, \frak t  \wedge  \frak t ,
\label{rlorentzian}
\end{equation}
with $\{\alpha,\beta,\gamma\}$ denoting generic tensor coefficients for each component of the $r$-matrix. Then it is straightforward to prove that the cocommutator~\eqref{cocomm} arising from~\eqref{rlorentzian} and the commutation rules~\eqref{aa} will be of the  form
\bea
&& \delta (\frak h) \subseteq \a  \, \frak h \wedge  \frak h \oplus \b  \, \frak h \wedge  \frak t  \oplus \c  \, \frak t  \wedge  \frak t ,\\
&& \delta (\frak t ) \subseteq \b  \, \Lambda\,\frak h \wedge  \frak h \oplus \a  \, \frak h \wedge  \frak t  \oplus \c  \, \Lambda\, \frak h \wedge  \frak t  \oplus  \b\, \frak t  \wedge  \frak t \, ,
\eea
where the cosmological constant parameter that distinguishes between the (A)dS and Poincar\'e cases appears explicitly.

After lengthy but straightforward computations involving explicitly the structure constants of the Lorentzian Lie algebras that we omit here for the sake of brevity, the following results can be proven:
\begin{enumerate}

\item A Lorentzian Lie bialgebra $(\mathfrak g_\Lambda,[\cdot,\cdot],\delta)$ is coisotropic iff $\c=0$.

\item For coisotropic Lie bialgebras ($\c=0$), coreductivity ($\delta(\frak t) \subseteq  \frak h \wedge \frak h \oplus \frak t \wedge \frak t 
$) is obtained iff $\a=0$. Therefore, coreductive Lorentzian Lie bialgebras are given exclusively by $r$-matrices of the form
\begin{equation}
r \subseteq \beta\,\frak h \wedge  \frak t ,
\label{rcosym}
\end{equation}
and whose coefficients are constrained by the modified CYBE.
This automatically precludes Lie bialgebras such that $\delta(\frak h ) \subseteq \frak h\wedge \frak h$ with $\delta(\frak h )\neq 0$ to be coreductive, and these are  Lie bialgebras associated to Poisson homogeneous spaces $M_\Lambda=G/H$ for which $H$ is a Poisson-Lie subgroup~\cite{BMN2017homogeneous}.

\item Cosymmetry for coreductive Lorentzian Lie bialgebras ($\delta(\frak h)$ does not contain $ \frak h \wedge  \frak h
$) is obtained  iff $\a=0$. Therefore, in this case if we have both coisotropy and coreductivity then cosymmetry is automatically verified. 

\item Note that the answer to the question whether coreductivity implies coisotropy is Lie algebra dependent: for $\Lambda = 0$ the answer is negative, while for $\Lambda \neq 0$ is a positive one. 

\item The same abovementioned conclusions can be extracted for Lie bialgebras corresponding to ${\frak {so}}(5)$ and  ${\frak {iso}}(4)$, where $\frak h={\frak {so}}(4)$,
since they are structurally equivalent to the Lorentzian ones with the cosmological constant being replaced by the inverse of the square of the radius of the sphere (see~\cite{BHOS1994global}).

\end{enumerate}

We recall that classifications of $r$-matrices for the (2+1) Poincar\'e and (A)dS Lie algebras have been presented, respectively, in~\cite{Stachura1998,BLT2016unified,BLT2016unifiedaddendum}, while for the (3+1) Poincar\'e case a classification can be found in~\cite{Zakrzewski1997}. Nevertheless, we remark that most of these results are not written in the kinematical basis~\eqref{lorschem}.
In the following we comment on some (2+1) and (3+1) Lorentzian examples, thus making more explicit the previous definition and constructions. As it has been already remarked, coisotropy and coreductivity  are quite restrictive properties and will select a very specific class of Lorentzian Poisson homogeneous spaces.


\subsection{Dual PHS and the $\kappa$-Lie bialgebra in (2+1) dimensions}

Let us firstly illustrate the previous construction of  coreductive Lie bialgebras and their complementary dual homogeneous spaces with the analysis of the Lie bialgebra associated to the well-known $\kappa$-deformation of the (2+1) dimensional Lorentzian algebras. This Lie bialgebra is coisotropic, coreductive and cosymmetric since it is generated by the $r$-matrix (see~\cite{LRNT1991,LNR1991realforms, BHOS1994global, ASS2004, BHM2014tallinn, BRH2017, BGGH2017curvedplb} and references therein) 
\be
r=z(K_1\wedge P_1+K_2\wedge P_2),
\label{ca}
\ee
which is indeed of the form~\eqref{rcosym} and where $z=1/\kappa$ is the `quantum' deformation parameter. We recall that 
in (2+1) dimensions the Lie brackets of  the Lie algebra $\frak g_\Lambda$ take the form
\be
\begin{array}{lll} 
\conm{J}{P_i}=   \epsilon_{ij}P_j , &\qquad
\conm{J}{K_i}=   \epsilon_{ij}K_j , &\qquad  \conm{J}{P_0}= 0  , \\[2pt]
\conm{P_i}{K_j}=-\delta_{ij}P_0 ,&\qquad \conm{P_0}{K_i}=-P_i ,&\qquad
\conm{K_1}{K_2}= -J   , \\[2pt]
\conm{P_0}{P_i}=-\Lambda\, K_i ,&\qquad \conm{P_1}{P_2}= \Lambda\, J  ,
\end{array}
\label{ba} 
\ee
where $i,j=1,2$, and  $\epsilon_{ij}$ is a skew-symmetric tensor with $\epsilon_{12}=1$. The (2+1) (A)dS and Minkowski spacetimes (that we will jointly denote as $M_\Lambda$) are obtained as $M_\Lambda=G_\Lambda/H$ where the isotropy subgroup $H$ is the (2+1) Lorentz group $SO(2,1)$ with Lie algebra $\frak h=\mbox{span}\{J, K_1, K_2 \}$. Therefore, $M_\Lambda$ turns out to be three-dimensional and is parametrized by the usual spacetime coordinates $\{x^0,x^1,x^2\}$, given by the composition of the canonical projection $p:G \rightarrow G/H, g \rightarrow gH$ with the coordinates on the Lie group $G$ defined by the inverse of the map
\begin{equation}
G=\exp{(x^0 P_0)} \exp{(x^1 P_1)} \exp{(x^2 P_2)} \exp{(\xi^1 K_1)} \exp{(\xi^2 K_2)} \exp{(\theta J)}.
\end{equation}
We stress that $\{x^0,x^1,x^2\}$ defined in this way are right invariant as functions on $G$.

The cocommutator map obtained from~\eqref{ca} reads
\begin{eqnarray}
&& \delta(P_0) =  \delta(J)=0 , \nonumber\\ 
&&  \delta (P_1)=   z (P_1\wedge P_0  + \Lambda \, K_2\wedge J  )  ,
\nonumber\\ 
&&
  \delta(P_2)=  z  (P_2\wedge P_0 - \Lambda\,  K_1 \wedge J)       ,
\label{cc}  \\
&&
  \delta(K_1)= z (K_1 \wedge P_0  + P_2 \wedge J)   ,
 \nonumber\\ 
&&
  \delta(K_2)=   z  (K_2  \wedge P_0 -P_1\wedge J)   ,
  \nonumber
\end{eqnarray}
which is indeed coisotropic with respect to the Lorentz Lie subalgebra $\frak h$.
Therefore, if we denote the dual generators to  $\{P_0,P_1,P_2,K_1,K_2,J\}$ by, respectively, $\{\hat x^0,\hat x^1,\hat x^2,\hat \xi^1,\hat \xi^2,\hat \theta\}$, the Lie brackets defining the (solvable) Lie algebra $\frak g^\ast$ of the dual Poisson-Lie group $G^\ast_\Lambda$ are straightforwardly deduced from~\eqref{cc} and read
\begin{equation} 
\begin{array}{lll} 
[\hat x^0, \hat x^1]_* =-z \, \hat x^1    ,
& \qquad [\hat x^0, \hat x^2]_* =-z \, \hat x^2  , 
& \qquad [\hat x^1, \hat x^2]_* =0
  , \\[2pt]
[\hat x^0, \hat \xi^1]_* =-z\,  \hat \xi^1    ,
&\qquad [\hat x^0, \hat \xi^2]_* =-z \, \hat \xi^2  , 
&\qquad [\hat \xi^1, \hat \xi^2]_* =0  ,
 \\[2pt]
[\hat \theta, \hat x^2]_* =-z \, \hat \xi^1  ,
 &\qquad[\hat \theta, \hat \xi^1]_* =z\,{\Lambda}\,  \hat x^2  , 
&\qquad [\hat \xi^1, \hat x^2]_* =0   ,
 \\[2pt]
[\hat \theta, \hat x^1]_* =z \, \hat \xi^2    ,
 &\qquad[\hat \theta, \hat \xi^2]_* =-z\,{\Lambda}\,  \hat x^1  , 
 &\qquad[\hat \xi^2, \hat x^1]_* =0  ,
 \\[2pt]
[\hat \theta, \hat x^0]_* =0    ,
&\qquad [\hat \xi^1, \hat x^1]_* =0  , 
 &\qquad[\hat \xi^2, \hat x^2]_* =0  .
\end{array}
\label{lie21}
\end{equation} 
On the other hand, the dual cocommutator map $\delta^*$ is obtained as the dual of the Lie bracket~\eqref{ba} for the $\frak g_\Lambda$ algebra, namely
\bea
&& \delta^*(\hat x^0) =   \hat \xi^1 \wedge \hat x^1 + \hat \xi^2 \wedge \hat x^2, \nonumber\\ 
&&  \delta^* (\hat x^1)=  - \hat \theta \wedge  \hat x^2 +  \hat \xi^1 \wedge  \hat x^0 ,\nonumber\\ 
&&  \delta^* (\hat x^2)=  \hat \theta \wedge \hat x^1 +  \hat \xi^2 \wedge  \hat x^0 ,\nonumber\\ 
&&  \delta^* (\hat \theta)= \Lambda\, \hat x^1 \wedge  \hat x^2 -  \hat \xi^1 \wedge \hat \xi^2 ,\label{dualads}\\ 
&&  \delta^* (\hat \xi^1)=  - \Lambda\,\hat \theta \wedge \hat \xi^2  ,\nonumber\\ 
&&  \delta^* (\hat \xi^2)=  \Lambda\,\hat \theta \wedge \hat \xi^1  .\nonumber
\eea
Note that $\Lambda$ plays now the role of a deformation parameter for the dual cocommutator, although in this case the $\Lambda \to 0$ limit does not lead to the zero cocommutator map.

We recall that the Poisson homogeneous Lorentzian spacetimes $(M_\Lambda,\pi)$ associated to the $\kappa$-PL structure given by~\eqref{ca} are explicitly given by the Poisson structure (see~\cite{BRH2017} for details)
\begin{equation} 
\begin{array}{l} 
\displaystyle{ \{x^0,x^1\}_\pi =-z\,\frac{\tan\sqrt{\Lambda} \,x^1}{ \sqrt{\Lambda}\,\cos^2\!(\sqrt{\Lambda} \,x^2)} , \qquad
\{x^0,x^2\}_\pi  =-z\,\frac{\tan(\sqrt{\Lambda} \, x_2)}{\sqrt{\Lambda}} ,\qquad 
\{x^1,x^2\}_\pi  =0 , } 
\end{array}
\label{gc}
\end{equation}  
which in the limit $\Lambda\to 0$ gives rise to the Poisson version of the well-known $\kappa$-Minkowski noncommutative spacetime
\begin{equation} 
\begin{array}{l} 
\displaystyle{ \{x^0,x^1\}_\pi  =-z\, x^1 , \qquad
\{x^0,x^2\}_\pi  =-z\,x^2  ,\qquad 
\{x^1,x^2\}_\pi  =0 . }
\end{array}
\label{gcp1}
\end{equation} 

Moreover, since the (A)dS and Minkowski spacetimes are obtained as cosets by the Lorentz isotropy subgroup generated by
$\frak h=\mbox{span}\{ J, K_1, K_2  \} $, we have that $\frak h^\perp=\mbox{span}\{\hat x^0,\hat x^1,\hat x^2\}$ and $\frak t^\perp = \mbox{span} \{\hat \xi^1,\hat \xi^2,\hat \theta\}$. Therefore, the commutation rules for $\frak g^\ast$~\eqref{lie21} are of the form
\begin{equation}
[\frak h^\perp,\frak h^\perp]_* \subseteq  \frak h^\perp 
\,, \qquad
[\frak h^\perp,\frak t^\perp]_* \subseteq   \frak t^\perp
\,, \qquad
[\frak t^\perp,\frak t^\perp]_* \subseteq  \Lambda\, \frak h^\perp 
\, .
\label{dualcocodS}
\end{equation} 
As a consequence, the reductive homogeneous spaces $M^\perp_\Lambda$ which are complementary dual to the Lorentzian spacetimes $M_\Lambda$ through the Lie bialgebra $\delta$, would be defined as $M^\perp_\Lambda=G^\ast_\Lambda/H^\perp$ where $H^\perp$ is the subgroup of $G^\ast$ generated by the dual translations $\{\hat x^0,\hat x^1,\hat x^2\}$. This is the so-called $\kappa$-Minkowski subgroup~\cite{Maslanka1993,MR1994,Zakrzewski1994poincare}, whose Lie algebra coincides with~\eqref{gcp1} (note that this algebra does not depend on $\Lambda$). In the limit $\Lambda\to 0$ (the Minkowski case) the $\hat \xi$ generators form an abelian subalgebra of infinitesimal translations on the dual space $M^\perp_0$. 

Also, we stress that $G^\ast_\Lambda$ with Lie algebra $(\frak g^\ast,[\cdot,\cdot]_*)$~\eqref{lie21} is by no means a semisimple Lie group, and the dual isotropy subgroup $H^\perp$ is a solvable one. Therefore, the associated Killing-Cartan form for the dual group $G^\ast$ is degenerate, and no guarantee of the existence of a $G^*$-invariant (indefinite) metric onto $M^\perp_\Lambda$ is expected. 
In fact, it is easy to prove by direct computation that for the particular case of the (2+1)-dimensional $\kappa$-Lie bialgebra, the dual space $M^\perp_\Lambda$ cannot be endowed with such a metric, just by noticing that $G^\ast$-invariant metrics on reductive spaces are in one-to-one correspondence with $\mathrm{ad}_{\mathfrak h^\perp}$-invariant non-degenerate symmetric bilinear forms $\langle \cdot , \cdot \rangle : \mathfrak{t}^\perp \times \mathfrak{t}^\perp \rightarrow \mathbb R$ on $\mathfrak{t}^\perp = \text{Lie}(T^\perp)$. So, in particular, the bilinear form satisfies 
\begin{equation}
\langle [Z,X]_*,Y\rangle + \langle X,[Z,Y]_* \rangle = 0, \qquad \forall X,Y \in \mathfrak{t}^\perp, Z \in \mathfrak{h}^\perp,
\end{equation}
where $\mathfrak{h}^\perp = \text{Lie}(H^\perp)$. This implies that, in general, the geometric features of the three-dimensional dual spaces $M^\perp_\Lambda$ will be quite different from their Lorentzian counterparts $M_\Lambda$.

Nevertheless, we stress that $M^\perp$ are Poisson homogeneous spaces whose Poisson structure $\pi^\ast$ can be also obtained as the canonical projection of the dual PL bracket $\Pi^\ast$ with associated Lie bialgebra $(\mathfrak g^\ast, [\cdot,\cdot]_*, \delta^*)$. The latter is, by construction, coisotropic with respect to the dual isotropy subalgebra generated by the generators of $\frak h^\perp$. Recall also that $\pi^\ast$ is defined on $\mathcal C^\infty(M^\perp) \times \mathcal C^\infty(M^\perp)$, and the coordinates on $M^\perp$ are the local coordinates associated to the $\hat \xi$ generators.

In the particular case of the $\kappa$-deformation,  the dual Lie bialgebra $(\mathfrak g^\ast, [\cdot,\cdot]_*, \delta^*)$ given by~\eqref{lie21} and~\eqref{dualads} is not a coboundary one since there is no $r$-matrix defined within $\frak g^\ast \otimes \frak g^\ast$ that could generate $\delta^*$. Despite of this fact (which implies that no Sklyanin bracket is available for the construction of $\Pi^\ast$) its full dual PL bracket $\Pi^\ast$ can be computed through the method based on a Poisson version of the quantum duality principle~\cite{Drinfeld1987icm, Semenov-Tyan-Shanskii1992, Gavarini2002, CG2006qd} which was introduced in~\cite{BM2013dual}  (see expressions~(11) in~\cite{BGGH2017curvedplb}). The canonical projection of this bracket onto the $\{K^\ast_1,K^\ast_2,J^\ast\}$ coordinates, which are the local coordinates associated   to the $\hat \xi=\{\hat \xi^1,\hat \xi^2, \hat \theta\}$ generators, respectively, gives the ${\pi^\ast}$ bracket for the dual PHS $M_\Lambda^\perp$, which reads
\be
\pois{J^\ast}{K_1^\ast}_{\pi^\ast} =     K_2^\ast    ,
\qquad   \pois{J^\ast}{K_2^\ast}_{\pi^\ast} =     -K_1^\ast   ,
\qquad   \pois{K_1^\ast}{K_2^\ast}_{\pi^\ast} =  -\frac{  \sin(2 z \sqrt{\Lambda} J^\ast)}{2z \sqrt{\Lambda} }.
\label{pldual}
\ee
It is worth emphasizing that, due to the non-coboundary nature of $(\mathfrak g^\ast, [\cdot,\cdot]_*, \delta^*)$, the construction of the PL bracket $\Pi^\ast$ on $G^\ast$ from which~\eqref{pldual} is obtained as a projection, was performed in~\cite{BGGH2017curvedplb} by imposing its Poisson map compatibility with the coalgebra structure provided by the group multiplication, and its computation is by no means a trivial one. For that reason, the coordinates employed to describe $G^\ast$ were fixed in such a way that they are well-defined functions on the coset space $M^\perp_\Lambda=G^\ast_\Lambda/H^\perp$, since the commutation rules of the dual generators of translations and boosts guarantee an appropriate ordering in the exponentiation of the dual group $G^\ast$ (see~Eq. (31) in~\cite{BGGH2017curvedplb}). We also stress that getting a common description in terms of `dual' coordinates of two different coset spaces and of their corresponding non-coboundary PHS is, in general, a difficult problem.

The dual Poisson homogeneous space $M^\perp_\Lambda=G^\ast/H^\perp$ endowed with~\eqref{pldual} deserves several comments:
\begin{itemize}

\item As expected, the linearization of~\eqref{pldual} coincides with the dual of the cocommutator $\delta^*$~\eqref{dualads}, or equivalently with the Poisson version of the commutation rules for the Lorentz Lie algebra sector in~\eqref{ba}.

\item When $z\neq 0$, the bracket~\eqref{pldual}  is a cosmological constant deformation of the $\mathfrak{so}(2,1)$ algebra, that is recovered in the limit $\Lambda\to 0$. This is similar to what occurs in~\eqref{gc}, which in the case $z\neq 0$ is just a $\Lambda$-deformation of the $\kappa$-Minkowski Lie algebra (see also~\cite{BRH2017}).

\item Therefore, we can say that the Poisson algebra $(M_\Lambda,\pi)$~\eqref{gc} is a cosmological constant deformation of the Lie algebra~\eqref{gcp1}, while the dual Poisson homogeneous space $(M^\perp_\Lambda,\pi^\ast)$~\eqref{pldual} is  a $\Lambda$-deformation of the Lie algebra generated by the dual coordinates to the translations in $M^\perp$, which is a Poisson analogue of the Lorentz subalgebra~\eqref{ba}.

\end{itemize}

Summarizing, we conclude that given any (2+1) dimensional Lorentzian Lie bialgebra, its complementary dual homogeneous space $M^\perp_\Lambda$ will have as its Poisson bracket $\pi^\ast$ either the Lorentz Lie algebra or a deformation of it.  As we will see in the following, all these are structural properties imposed by the coreductivity condition, and will also appear in (3+1) dimensions.


\subsection{The (3+1) dimensional case}

In (3+1) dimensions the $r$-matrix for the $\kappa$-deformation of Lorentzian Lie algebras  is given by (see~\cite{BGH2019kappaAdS3+1,BGHOS1995quasiorthogonal,BHMN2017kappa3+1})
\be
r=z \left( K_1 \wedge P_1 + K_2 \wedge P_2+ K_3 \wedge P_3 + \sqrt{-\Lambda}\, J_1 \wedge J_2 \right),
\label{ab}
\ee
which is always coisotropic with respect to the Lorentz subalgebra. However, due to the presence of the $ \sqrt{-\Lambda}\, J_1 \wedge J_2 $ term, only for $\Lambda=0$ this $r$-matrix gives rise to a coreductive (and thus cosymmetric) Lie bialgebra structure. It should be noted here that the parameter $z=1/\kappa$ does not contain any mathematical information in this respect. However, we choose to keep it explicitly in order to make contact with the physics literature, since in this way $z$ has dimensions of time.
In particular, the cocommutator derived from~\eqref{ab} is
 \begin{align}
\delta(P_0)&=0,\qquad \delta(J_3)=0 , \nonumber \\
\delta(J_1)&=z \sqrt{-\Lambda}   J_1 \wedge J_3  ,\qquad 
\delta(J_2)=z \sqrt{-\Lambda}   J_2 \wedge J_3   ,\nonumber \\
\delta(P_1)&=z \left(P_1 \wedge P_0 +\Lambda   J_2  \wedge  K_3 -\Lambda J_3 \wedge K_2 + \sqrt{-\Lambda}J_1 \wedge P_3 \right) , \nonumber\\
\delta(P_2)&=z \left(P_2 \wedge P_0 +\Lambda  J_3  \wedge K_1-\Lambda J_1 \wedge K_3 + \sqrt{-\Lambda}  J_2 \wedge P_3 \right) ,\nonumber\\
\delta(P_3)&=z \left(  P_3 \wedge P_0+\Lambda J_1\wedge  K_2  -\Lambda J_2 \wedge K_1  - \sqrt{-\Lambda} J_1  \wedge P_1 - \sqrt{-\Lambda}  J_2  \wedge P_2 \right), \label{ac}\\
\delta(K_1)&=z \left(K_1 \wedge P_0 + J_2 \wedge P_3 - J_3 \wedge  P_2  +\sqrt{-\Lambda}  J_1 \wedge K_3  \right), \nonumber\\
\delta(K_2)&=z \left(   K_2 \wedge P_0+ J_3 \wedge P_1 - J_1  \wedge P_3 +\sqrt{-\Lambda} J_2 \wedge K_3   \right) ,\nonumber \\
\delta(K_3)&=z \left(  K_3 \wedge P_0 + J_1 \wedge P_2 - J_2  \wedge P_1  - \sqrt{-\Lambda} J_1   \wedge K_1  -\sqrt{-\Lambda}  J_2  \wedge  K_2 \right)  .
\nonumber
\end{align}

This means that reductive complementary duals of (A)dS spaces are excluded in (3+1) dimensions. On the contrary, the dual $(M^\perp_0,\pi^\ast)$ of the $\kappa$-Minkowski Poisson homogeneous space $(M_0,\pi)$ can be constructed as a reductive space. This $M^\perp_0$ space will be 6-dimensional, since $G^\ast$ is 10-dimensional and the isotropy subgroup for $M^\perp_0$ will be generated by the $\{\hat x^0,\hat x^1,\hat x^2,\hat x^3\}$ generators dual to the $P_\mu$ translations. 

In this case the $\kappa$-Minkowski~\cite{Maslanka1993, MR1994, Zakrzewski1994poincare} Poisson homogeneous spacetime $(M_0,\pi)$ obtained by projecting the Sklyanin bracket for the $r$-matrix~\eqref{ab} onto the spacetime coordinates can be proven to be given (see, for instance,~\cite{BGH2019worldlinesplb}) by the (3+1)-dim\-ensional generalization of~\eqref{gcp1}, namely
 \be  
\begin{array}{l} 
\displaystyle{ \{x^0,x^1\}_\pi  =-z\, x^1 , \quad
\{x^0,x^2\}_\pi  =-z\,x^2  ,\quad 
\{x^0,x^3\}_\pi  =-z\,x^3  ,\quad 
\{x^i,x^j\}_\pi  =0 , \quad i,j=1,2,3. }
\end{array}
\label{gcp}
\ee 
On the other hand, the Poisson homogeneous structure $(M^\perp_0,\pi^\ast)$ has to be obtained as the canonical projection onto the $\xi^\ast$ sector of the corresponding PL structure on $G^\ast$, which was explicitly constructed in~\cite{BHMN2017kappa3+1} by following the method introduced in~\cite{BM2013dual}. It is straightforward to check that this projection gives rise to a Poisson structure $\pi^\ast$ which is a (undeformed) Poisson version of the Lorentz Lie algebra, namely 
\be
\label{eq:dualpoissonmink3+1}
\pois{J_a^\ast}{J_b^\ast}_{\pi^\ast}=\epsilon_{abc}J_c^\ast , 
\qquad \pois{J_a^\ast}{K_b^\ast}_{\pi^\ast}=\epsilon_{abc}K_c^\ast , 
\qquad
\pois{K_a^\ast}{K_b^\ast}_{\pi^\ast}=-\epsilon_{abc} J_c^\ast \, ,
\ee
which again generalizes the $\Lambda\to 0$ case of the (2+1)-dimensional construction~\eqref{pldual}. As it was mentioned in the previous section, this Poisson homogeneous structure $(M^\perp_0,\pi^\ast)$ is well defined because $\{J_a^\ast,K_a^\ast\}$ are, by construction, suitable coordinates on the coset space $M^\perp_0$. This statement can be explicitly checked from in equation~(19) in~\cite{BGGH2018cms31}, where the PL structure on $G^\ast$ was obtained, since the vanishing commutation relations among dual generators of boosts and translations ensures that the ordering in the exponentiation is the suitable one for the description of the coset space in terms of local coordinates. As a consequence, the noncommutative spacetime arising from quantizing $\pi^\ast$ will be just isomorphic to the Lorentz Lie algebra $\frak{so}(3,1)$, whose representation theory is well-known~\cite{Wu-KiTung1985book}.


\section{On the geometry of dual Poisson homogeneous spaces}

In the previous Section we have dealt with the Poisson geometry of some spaces $M^\perp$ since, by construction, these spaces are naturally endowed with a Poisson structure compatible with the left action of $G^*$. The general method for constructing such Poisson homogeneous structure on $M^\perp$ has been given, and some interesting examples have been worked out in detail. However, while on the spacetimes $M_\Lambda$ the pseudo-Riemannian structure coexists with the Poisson structure in a natural way, the first one describing the classical geometry (general relativity) and the second one describing semi-classical quantum corrections, we have seen that, in general, the dual PHS $M^\perp_\Lambda$ do not admit  a $G$-invariant pseudo-riemannian metric. Therefore, alternative approaches for the characterization of the geometric properties of the dual PHS are needed.

A natural approach is to consider the general setting of $K$-structures on manifolds, by following~\cite{chern1953Gstructures} and~\cite{KobayashiNomizu1969-2}. Let $M^\perp$ be the PHS dual to a coreductive Lie bialgebra $(\mathfrak g, [\cdot,\cdot], \delta)$, and let $w$ be the dimension of $M^\perp$. Consider the frame bundle $F(M^\perp)$ viewed as a principal bundle over $M^\perp$ with structure group $GL(w,\mathbb{R})$. With this notation, a $K$-structure is a reduction of $F(M^\perp)$ to the subgroup $K$ of $GL(w,\mathbb{R})$. A connection in the principal bundle defined by the $K$-structure on $M^\perp$ induces a linear connection on the tangent bundle of the manifold $M^\perp$ which is said to be adapted to the $K$-structure. Associated to each connection we have its torsion and curvature, which indeed give information not only about the geometry but also the topology of the manifold, and this will be the route we propose in order to extract some explicit geometric information about the space $M^\perp$.

We recall that the torsion and curvature tensors $T$ and $R$ of a given connection are defined~\cite{KobayashiNomizu1963-1} in terms of the covariant differentiation $\nabla$ associated to it, namely
\begin{equation}
\begin{split}
&T(X,Y) =\nabla_X Y - \nabla_Y X - [X,Y], \\
&R(X,Y)Z = \left[ \nabla_X, \nabla_Y \right] Z - \nabla_{[X,Y]} Z, \qquad \forall X,Y,Z \in \mathfrak{X} (M^\perp).
\end{split}
\end{equation}
In terms of the linear map 
\begin{equation}
\begin{split}
\varphi : \; & T_w M^\perp \rightarrow T_w M^\perp \\
&X \rightarrow R(X,Y) Z\qquad \forall\,X,Y,Z \in T_w M^\perp,
\end{split}
\end{equation}
the Ricci tensor $S$ is defined by 
\be
S(Y,Z)=\mbox{tr} \; \varphi,
\qquad
\forall\,Y,Z  \in T_w M^\perp.
\ee
It should be stressed that such connections are far from being unique (so having different associated torsion and curvature forms), but in the particular case of reductive spaces the so-called canonical connection having a particularly simple form can be defined. 

Let us consider the dual PHS corresponding to the coreductive Lie bialgebra $(\mathfrak{g}, [\cdot,\cdot], \delta)$ defined as the coset space $M^\perp=G^*/H^\perp$, and let us denote by $eH^\perp$ to its origin. We know that $\mathfrak{g}^* = \text{Lie} (G^*)$ admits an $\mathrm{Ad}_{\mathfrak h}$-invariant splitting of the form $\mathfrak{g}^* = \mathfrak{t}^\perp \oplus \mathfrak{h}^\perp$ where $ \mathfrak{h}^\perp = \text{Lie} (H^\perp)$, so we can identify $T_{eH^\perp} M^\perp \simeq \mathfrak{t}^\perp$. We define the Lie bracket projection onto the subspaces associated to this decomposition as 
\begin{equation}
\left[ X,Y \right]_* = \left[ X,Y \right]_{*,\mathfrak{t}^\perp} + \left[ X,Y \right]_{*,\mathfrak{h}^\perp}, \qquad \forall X,Y \in \mathfrak{g}^*,
\end{equation}
where $ \left[ \cdot,\cdot  \right]_{*,\mathfrak{t}^\perp}$ stands for the projection to the subspace $\mathfrak{t}^\perp$ of the Lie bracket $ \left[ \cdot,\cdot \right]_*$ on $\mathfrak{g}^*$. With this notation, the so-called canonical connection for the dual PHS corresponding to a coreductive Lie bialgebra fulfills the following relations~\cite{KobayashiNomizu1969-2} for all $ X,Y,Z \in \mathfrak{t}^\perp$: 
\begin{equation}
\begin{split}
&T(X,Y)_{eH^\perp} = - \left[ X,Y \right]_{*,\mathfrak{t}^\perp}, \\
&\left(R(X,Y)Z \right)_{eH^\perp} = -  \left[  \left[ X,Y \right]_{*,\mathfrak{h}^\perp},Z \right]_*, \\
&\nabla\,T = 0, \\
&\nabla\, R = 0.
\end{split}
\end{equation} 
The last two identities are a direct consequence of the fact that every $G^*$-invariant tensor field is parallel transported by the canonical connection. Here it should be noticed that the canonical connection just defined is complete for every complementary dual PHS corresponding to a coreductive Lie bialgebra, and that its set of geodesics passing through the origin is given by $\{ (\exp t X).eH^\perp \, | \, X \in \mathfrak t^\perp \}$.

In the case of $M^\perp$ being the dual PHS corresponding to a cosymmetric Lie bialgebra $(\mathfrak{g}, [\cdot,\cdot], \delta)$, further simplifications arise for the torsion and curvature tensors by taking into account that $ \left[ \mathfrak{t}^\perp,\mathfrak{t}^\perp \right]_* \subseteq  \mathfrak{h}^\perp$, and therefore 
\begin{equation}
\begin{split}
&T(X,Y)_{eH^\perp} = 0, \\
&\left(R(X,Y)Z \right)_{eH^\perp} = -  \left[  \left[ X,Y \right]_*,Z \right]_*, \\
&\nabla T = 0, \\
&\nabla R = 0,
\end{split}
\end{equation}
for all $ X,Y,Z \in \mathfrak{t}^\perp$. In general, the fact that the torsion tensor vanishes identically if $(\mathfrak{g}, [\cdot,\cdot], \delta)$ is cosymmetric means that the canonical connection on $M^\perp$ coincides with the Levi-Civita connection associated to a $G^*$-invariant Riemannian metric (provided it exists).

As an example, let $(\mathfrak{g}, [\cdot,\cdot], \delta)$ be the $\kappa$-Lie bialgebra in (2+1) dimensions given by \eqref{ba} and \eqref{cc}. A straightforward computation shows that the only non vanishing components of the curvature tensor $R$ are 
\begin{equation}
\left(R(\hat{\xi}^i,\hat{\theta}),\hat{\theta}\right)_{eH^\perp} = z^2 \Lambda \, \hat{\xi}^i, \qquad i\in \{1,2\}.
\end{equation}
while the Ricci tensor $S$ has the only non-vanishing component given by
\begin{equation}
\left(S(\hat{\theta},\hat{\theta})\right)_{eH^\perp} = 2 z^2 \Lambda
\end{equation}
Therefore, the complementary dual space $M^\perp$ associated to the $\kappa$-Lie bialgebra in (2+1) dimensions turns out to be Ricci flat iff the corresponding model spacetime $M_\Lambda$ is flat, i.e. only in the Minkowski case where $\Lambda=0$ (we recall that the limit $z\to 0$ corresponds to the trivial PHS structure with Abelian dual group $G^\ast$). At this stage  the scalar curvature of $M^\perp$ cannot be defined because we have not endowed it with a metric. Note that although we have previously proved that $M^\perp$ does not admit a $G^*$-invariant metric, such $G^*$-invariance condition -which in fact is quite restrictive- could perhaps be relaxed.

For the $(3+1)$ dimensional $\kappa$-Poincar\'e deformation, whose dual Lie algebra was presented in \cite{BGGH2018cms31}, it is straightforward to check that $\mathfrak t^\perp$ is a commutative Lie subalgebra, so the Riemann tensor for the dual space $M^\perp_0$, with Poisson structure given in \eqref{eq:dualpoissonmink3+1}, vanishes identically.


\section{Coreductivity and uncertainty relations}

The main physical motivation for the introduction of noncommutative spacetimes relies on the widely shared idea that some quantum gravity effects could be described (in an effective or dynamical way) by  a suitable `quantum' geometry in which spacetime coordinates are replaced by noncommutative operators (see for instance~\cite{Maggiore1993algebraicgup, Szabo2003, PW1990, BM2018extended, Snyder1947, DFR1995, CKNT2004} and references therein). In particular, quantum homogeneous spaces $M_q$ provide instances of such noncommutative spacetimes for which the notion of covariance under the corresponding quantum groups $G_q$ can be implemented. In this context the dual Lie algebra $(\frak g^\ast,[\cdot,\cdot]_*)$ provides  the first order of the noncommutative algebra defining $G_q$, while the first order of the noncommutative spacetime's commutation relations $M_q$ is given by the $\frak h^\perp$ subalgebra, generated by the $\hat x$ operators. Therefore, the noncommutativity of the algebra of coordinates of spacetime events implies the existence of Heisenberg-type uncertainty relations in the case of simultaneous measurements of different components of the noncommutative coordinates $\hat x$ (and their functions).

As we have recalled, by definition of a coisotropic PHS, the coisotropy condition~\eqref{coisotropy} for a given Lie bialgebra with respect to a fixed Lie subalgebra guarantees that the `quantum' spacetime coordinates $\hat x$ close a subalgebra within the dual Lie algebra $(\frak g^\ast,[\cdot,\cdot]_*)$ of quantum group coordinates, and that $\frak h^\perp$ generates the isotropy subgroup of the complementary dual homogeneous space $M^\perp$. Moreover, by definition, the coreductivity condition~\eqref{coredt} imposes onto $(\frak g^\ast,[\cdot,\cdot]_*)$ the condition of being the Lie algebra of a reductive space (of course when the coisotropy condition is also satisfied). This fact will be reflected in the representation theory of $(\frak g^\ast,[\cdot,\cdot]_*)$, a fact which has far-reaching consequences from a physical viewpoint.

Let us firstly assume that the dual Lie algebra $(\mathfrak g^*, [\cdot,\cdot]_*)$ can be endowed with a $C^\ast$-algebra structure, and let us consider a unitary irreducible representation of this algebra on a Hilbert space of physical states denoted by $|\psi\rangle$. Then, if coreductivity does not hold this means that we allow for elements of $\hat x$ to appear on the right-hand-side of the commutation rules $[\hat x,\hat \xi]_* $ in the dual Lie algebra, namely,
\begin{equation}
[\hat x,\hat x]_* \subseteq \hat x 
\,, \qquad
[\hat x,\hat \xi]_* \subseteq  \hat \xi + \hat x
\,, \qquad
[\hat \xi,\hat \xi]_* \subseteq  \hat \xi + \hat x \, .
\label{dualcocodSncr}
\end{equation} 
In that case there will exist at least one uncertainty relation of the form
\begin{equation}\label{Uncertainty_Relations_non_coreductive}
\Delta \hat x \, \Delta \hat \xi \geq {\frac 1 2} \langle \hat \xi \rangle  + {\frac 1 2} \langle \hat x \rangle \,,
\qquad\qquad
\text{where} \qquad \Delta \hat y = \sqrt{\langle \hat y^2 \rangle - \langle \hat y \rangle^2} \,.
\end{equation}
Now, let us consider the subset of states such that $\hat \xi |\psi\rangle = 0$. Since by definition such states have vanishing uncertainty  and $\Delta \hat\xi = \langle \psi | \hat \xi ^2 | \psi \rangle - \langle \hat \xi  \rangle^2=0$, then relations~(\ref{Uncertainty_Relations_non_coreductive}) impose singular constraints onto the expectation values of the momenta of $\hat x$. In particular,~\eqref{Uncertainty_Relations_non_coreductive} implies that either $\langle \psi | \hat x  | \psi \rangle =0$ or $\Delta \hat x \to \infty$. If we consider the representation space for the Lie subalgebra $[\hat x,\hat x]_* \subseteq  \hat x$ alone, there could be some states such that $\langle \hat x \rangle =0$, but these ones most certainly do not exhaust, in general, the set of all possible states. Similarly, there could  be sequences of states for the subalgebra generated by $\hat x$ whose uncertainty is divergent, but, again, they will not be generic ones.

On the contrary, if the coreductivity condition holds then we always have commutation rules of the type
\be
[\hat x,\hat \xi]_* \subseteq   \hat \xi ,
\label{redireps}
\ee
which give rise to uncertainty relations of the form
\begin{equation}
\Delta \hat x \, \Delta \hat \xi \geq {\frac 1 2} \langle \hat \xi \rangle \,.
\label{crossed}
\end{equation}
Now, if we consider the set of eigenstates of $\hat \xi$ with vanishing eigenvalue, $\hat \xi |\psi\rangle = 0$, then on these states the  `crossed' uncertainty relations~\eqref{crossed} do not constrain in any way the momenta of the spacetime observables $\hat x$. This argument can be illustrated with a well-known example: the theory of unitary irreducible representations (UIR) of the Poincar\'e Lie algebra. As it is explained in~\cite{Wu-KiTung1985book} (Section 10.4) the UIR of null-vector type are the ones with zero eigenvalues for the generators of the subalgebra of translations (the $\hat\xi $ operators in~\eqref{redireps}), and in this case the representation theory for the isotropy subgroup (the $\hat x$ operators) completely decouples with the one for the translation generators.

In summary, if there are $\hat x$ contributions on the right-hand-side of $[\hat x ,\hat \xi]_*$, the subset of  states for the $(\frak g^\ast,[\cdot,\cdot]_*)$ Lie algebra such that $\hat \xi |\psi\rangle = 0$ cannot provide with the full set of representation states for the noncommutative spacetime subalgebra $\hat x$. Therefore, the coreductivity condition allows us to get a physical insight of the subalgebra $\hat x$ on its own, and consider it as the keystone for the construction of the full noncommutative algebra of functions on a quantum homogeneous space $M_q$.

Another specific illustration of this argument can be extracted from the recent work~\cite{LMMP2018localization}, where the representation theory for the $\kappa$-Minkowski spacetime has been thouroughly studied. In the simpler (1+1)-dimensional case, the commutation relations between the noncommutative coordinates over the full quantum (1+1) Poincar\'e group are (see also~\cite{BHOS1995jmp})
\begin{equation}\label{1+1_kappaPoincareGroup_v3}
\begin{gathered}
{}[ \hat a^0 , \hat a^1 ]_* =  i \lambda \, \hat a^1  \,,
\qquad
[  \hat\xi , \hat a^0 ]_* = - i \lambda \sinh \hat \xi \,,
\qquad
[\hat \xi , \hat a^1 ]_* =  i \lambda  \left( 1- \cosh \hat \xi   \right) \, .
\end{gathered}
\end{equation}
Note that the linearization of these relations leads to
\begin{equation}\label{1+1_kappaPoincareGrouplinear_v3}
\begin{gathered}
{}[ \hat a^0 , \hat a^1 ]_* =  i \lambda \, \hat a^1  \,,
\qquad
[  \hat\xi , \hat a^0 ]_* = - i \lambda \,\hat \xi \,,
\qquad
[\hat \xi , \hat a^1 ] _*= 0 \, ,
\end{gathered}
\end{equation}
which exactly coincides with the (1+1)-dimensional version of~\eqref{lie21} in the case $\Lambda=0$, provided that $z=-i\lambda$ and under the identification of the quantum group translation coordinates as $\hat x^0 = \hat a^0$ and $\hat x^1 = \hat a^1$. Again,~\eqref{1+1_kappaPoincareGrouplinear_v3} illustrates the fact that the dual Lie algebra $(\mathfrak g^\ast, [\cdot,\cdot]_*)$~\eqref{lie21} is just the linearization of the full quantum group relations~\eqref{1+1_kappaPoincareGroup_v3}.

As is explicitly shown in~\cite{LMMP2018localization}, finite translation ($\hat a^\mu$) and Lorentz rapidity ($\hat \xi$) operators can be represented as differential operators on a Hilbert space of functions  on the Cartesian product between the Lorentz group and $\mathbbm{R}$ in the following way:
\begin{eqnarray}
\label{final_1+1_k-Poinc_representation}
\hat a^0  &=&  i \lambda \left(\frac 1 2 + q \frac{\partial}{\partial q} \right)  + i \lambda \left(  \frac 1 2 \cosh \xi  + \sinh \xi \, \frac{\partial}{\partial \xi}\right) \,, \nonumber\\
\hat a^1 &=& q + i \lambda \left( \frac 1 2 \sinh \xi +  \left( \cosh \xi  - 1 \right) \, \frac{\partial}{\partial \xi}   \right)  \,,
\end{eqnarray}
while $\xi$ is the coordinate associated with the eigenvalues of the multiplicative operator $\hat \xi$. The meaning of our uncertainty-relation argument is made clear in~\cite{LMMP2018localization}, where it is shown that there exists a sequence of well-normalized product wavefunctions $Q(\xi)$, such that for any function $f(q)$, all the expectation values
\begin{equation}
\langle Q(\xi)f(q) |  (\hat a^1)^n (\hat a^0)^m | Q(\xi)f(q) \rangle 
\, ,
\end{equation}
tend to the following:
\begin{equation}
\langle f(q) |  (\hat y^1)^n (\hat y^0)^m | f(q) \rangle 
\, ,
\end{equation}
where now $\hat y^\mu$ provide a faithful representation of the commutation relations of the $\kappa$-Minkowski quantum homogeneous space given by:
\begin{eqnarray}
\label{final_1+1_k-Mink_representation}
\hat y^0  &=&  i \lambda \left(\frac 1 2 + q \frac{\partial}{\partial q} \right)   \,, \nonumber\\
\hat y^1 &=& q  \,.
\end{eqnarray}
In this way we see that, by choosing a product state between this sequence of functions $Q(\xi)$ (which tend, in an appropriate way, to a function localized at $\xi=0$) and an arbitrary wavefunction $f$, we can reproduce the expectation values of any polynomial in $\hat y^0$ and $\hat y^1$, and so we can define the whole wealth of possible states on the $\kappa$-Minkowski algebra as a limit of states on the $\kappa$-Poincar\'e group, in which the $\xi$ contribution is sent to zero in a controlled way.


\section{Concluding remarks}

Summarizing, given a pointed Poisson homogeneous space $(M,\pi)$, where $M=G/H$ and the Poisson-Lie structure $\Pi$ on $G$  is characterized by the Lie bialgebra $(\mathfrak g, [\cdot,\cdot], \delta)$, the coisotropy, coreductivity and cosymmetry conditions for $\delta$ with respect to $\mathfrak h$ are given as the following constraints \begin{equation}\label{Cocommutators_3}
\begin{gathered}
\delta(\frak h) \subseteq  \cancel{\frak h \wedge \frak h}^\text{~cosymmetry} \oplus \frak h \wedge \frak t \oplus \cancel{\frak t \wedge \frak t}^\text{~coisotropy} \,,
\\
\delta(\frak t) \subseteq  \frak h \wedge \frak h \oplus \cancel{\frak h \wedge \frak t}^\text{~coreductivity} \oplus \frak t \wedge \frak t \,,
\end{gathered}
\end{equation}
thus leading to a dual Lie algebra $(\frak g^\ast, [\cdot,\cdot]_*)$ which is $(\mathfrak{h}^\perp, \mathfrak t^\perp)$-reductive and symmetric
\begin{equation}\label{Symmetric_space_conditions_2b}
[\frak h^\perp,\frak h^\perp]_* \subseteq \frak h^\perp \oplus \bcancel{\frak t^\perp}_\text{~coisotropy}
\,, \quad
[\frak h^\perp,\frak t^\perp]_* \subseteq \bcancel{\frak h^\perp}_\text{~coreductivity} \oplus \frak t^\perp
\,, \quad
[\frak t^\perp,\frak t^\perp]_* \subseteq \frak h^\perp \oplus \bcancel{\frak t^\perp}_\text{~cosymmetry}
\,.
\end{equation} 
When all these conditions are fulfilled, the complementary dual Poisson homogeneous space $(M^\perp=G^\ast/H^\perp,\pi^\ast)$ to $(M,\pi)$ is a reductive and symmetric homogeneous space for $G^*$. This construction is self-dual and so, the complementary dual Poisson homogeneous space to $(M^\perp=G^\ast/H^\perp,\pi^\ast)$ is $(M=G/H,\pi)$.

This self-dual picture has been fully illustrated by considering the well-known example of the Poincar\'e group $G$, its associated Minkowski spacetime $M_0=G/H$, a coisotropic and coreductive (with respect to the Lorentz subgroup $H$) PL  structure on $G$ (for instance, the one provided by the $\kappa$-deformation), together with their PL dual group $G^\ast$ and the dual space $M_0^\perp$. The four Poisson structures and their relations arising under such a duality picture are the following:

\begin{enumerate}

\item $\Pi$: The PL structure on the Poincar\'e group given by the $r$-matrix which corresponds to the $\kappa$-deformation. The linearization of this PL bracket is in one-to-one correspondence with the Lie bialgebra $(\mathfrak g, [\cdot,\cdot], \delta)$.

\item $\pi$: The Poisson homogeneous structure on the Minkowski spacetime $M_0=G/H$ (the Poisson $\kappa$-Minkowski spacetime), whose bracket can be obtained through canonical projection from $\Pi$, since $\delta$ is coisotropic with respect to $\frak h=\mbox{Lie}(H)$.

\item $\Pi^*$: The PL structure on the dual Poincar\'e group $G^\ast$, whose associated Lie bialgebra is $(\frak g^\ast, [\cdot,\cdot]_*, \delta^*)$. As we have shown, the Killing-Cartan form for $G^\ast$ is degenerate ($G^\ast$ is a solvable Lie group), and thus no $G^\ast$-invariant metric exists. 

\item $\pi^*$: The Poisson homogeneous structure on the dual spacetime $M_0^\perp$. Since the Lie bialgebra $(\mathfrak g, [\cdot,\cdot], \delta)$ is coreductive, the dual reductive homogeneous space $M_0^\perp=G^\ast/H^\perp$ can be defined. The dual Lie bialgebra $(\frak g^\ast, [\cdot,\cdot]_*, \delta^*)$ is also coisotropic. Thus, the bracket $\pi^*$ can be obtained through canonical projection from $\Pi^\ast$, and it provides a Poisson structure on the Lorentz group coordinates. Despite that $M_0^\perp$ cannot be endowed with a $G^\ast$-invariant metric, its geometry can be analysed from the viewpoint of $K$-structures and turns out to be torsionless and with vanishing curvature tensor.
\end{enumerate}

We would like to stress that the coreductivity and cosymmetry conditions provide a novel insight into the structural properties of Lie bialgebras (and therefore, of PL groups and PHS) which -to the best of our knowledge- had not been considered yet. In particular, as far as Lorentzian Lie bialgebras are concerned, we have shown that the coreductivity condition (with respect to the Lorentz subalgebra $\mathfrak h$ of the stabilizer of the origin) imposes very strong constraints: in general, non-trivial Poisson-subgroup homogeneous spacetimes are precluded, and the (A)dS cases with non-vanishing cosmological constant also face strong obstructions in the (3+1) dimensional case. In particular, the $\kappa$-Poincar\'e Lie bialgebra is coreductive (with respect to the Lorentz subalgebra $\mathfrak h$ of the stabilizer of the origin) in any dimension, while the $\kappa$-(A)dS Lie bialgebra is only coreductive (with respect to $\mathfrak h$) in (2+1) dimensions.

Finally, we recall that the corresponding quantum homogeneous spacetimes $M_q$ and $M_q^\perp$ will be just the quantizations of the Poisson homogeneous spaces $(M,\pi)$ and $(M^\perp,\pi^\ast)$, respectively. From a physical perspective, we have seen that the coreductivity condition for $\delta$ guarantees that the representation theory of the noncommutative spacetime algebra $M_q$ obtained by quantizing $(M,\pi)$  leads to uncertainty relations that are consistent with the notion of quantum group invariance, in the sense that the uncertainty relations arising from the commutation rules between the noncommuting coordinates on the full quantum group $G_q$ will be such that they admit sequences of states on $G_q$ whose limit behave exactly like the states on $M_q$ taken alone. In other words, the states on $M_q$ are equivalent to states on $G_q$ for which the quantum isotropy subgroup operators are localized on the identity transformation with vanishing uncertainty. By construction, the same would happen with the representation theory on the quantum analogue $M_q^\perp$ of the dual space $(M^\perp,\pi^\ast)$.


\section*{Acknowledgements}

This work has been partially supported by Ministerio de Ciencia e Innovaci\'on (Spain) under grants MTM2016-79639-P (AEI/FEDER, UE) and PID2019 - 106802GB-I00 / AEI / 10.13039 / 501100011033, by Junta de Castilla y Le\'on (Spain) under grants BU229P18 and GIR2019, as well as by the Action CA18108 QG-MM from the European Cooperation in Science and Technology (COST). 


\small


\end{document}